\documentstyle[epsfig,graphics]{mn}

\begin{document}

\newcommand{\Zsolar}{\mbox{\,$\rm Z_{\odot}$}}
\newcommand{\etal}{{et al.}\ }
\newcommand{\ang}{\mbox{$\rm \AA$}}
\newcommand{\xs}{$\chi^{2}$}
\newcommand{\be}{\begin{equation}}   
\newcommand{\ee}{\end{equation}}     
\def\japsub{\scriptscriptstyle \rm}
\def\kms{{\,\rm km\,s^{-1}}}

\title[The host galaxy of HDF850.1]{Discovery of the host galaxy of
HDF850.1, the brightest sub--mm source in the Hubble Deep Field}

\author[J.S. Dunlop et al..]
{J.S. Dunlop$^{1}$, R.J. McLure$^{1}$, T. Yamada$^{2}$, 
M. Kajisawa$^{2}$, J.A. Peacock$^{1}$, R.G. Mann$^{1}$, \and 
D.H. Hughes$^{3}$, I. Aretxaga$^{3}$,  T.W.B. Muxlow$^{4}$, 
A.M.S. Richards$^{4}$,
M. Dickinson$^{5}$, \and R.J. Ivison$^{6}$, G.P. Smith$^{7}$, I. Smail$^{7}$,
S. Serjeant$^{8}$, O. Almaini$^{1}$ \& A. Lawrence$^{1}$.
\\
$^{1}$Institute for Astronomy, University of Edinburgh, Royal Observatory,
Blackford Hill, Edinburgh, EH9 3HJ, UK\\
$^{2}$National Astronomical Observatory, 2-21-1, Osawa, Mitaka, Tokyo 181-8588, Japan\\
$^{3}$Instituto Nacional de Astrof\'{\i}sica, \'{O}ptica y Electr\'{o}nica
(INAOE), Apartado Postal 51 y 216, 72000 Puebla, Pue., Mexico\\
$^{4}$University of Manchester, MERLIN/VLBI National Facility, Jodrell Bank 
Observatory, Cheshire, SK11 9DL, UK\\
$^{5}$Space Telescope Science Institute, Baltimore MD21218, USA\\ 
$^{6}$UK ATC, Royal Observatory, Blackford Hill, Edinburgh, EH9 3HJ, UK\\
$^{7}$Department of Physics, University of Durham, South Road, Durham DH1 
3LE, UK\\
$^{8}$Centre for Astrophysics \& Planetary Science, School of Physical 
Sciences, University of Kent, Canterbury CT2 7NZ, UK}

\date{Submitted for publication in MNRAS}

\maketitle
  
\begin{abstract}
Despite extensive observational efforts, the brightest sub--mm source in the 
Hubble Deep Field, HDF850.1, has failed to yield a convincing optical/infrared 
identification almost 4 years after its discovery. This failure is all the 
more notable given the availability of supporting multi-frequency data of 
unparalleled depth, and sub-arcsec positional accuracy 
for the sub-mm/mm source. Consequently, HDF850.1 has become a test
case of the possibility that the most violently star-forming objects in the 
universe are too red and/or distant to be seen in the deepest optical images.

Here we report the discovery of the host galaxy of HDF850.1. This object
has been revealed by 
careful analysis of a new, deep $K^{\prime}$ image of the HDF obtained with the
Subaru 8.2-m telescope. Its reality is confirmed by a similar analysis of 
the HST NICMOS F160W image of the same region.
This object is extremely faint ($K\simeq 23.5$),
clumpy (on sub-arcsec scales) and very red
($I-K>5.2$; $H-K=1.4 \pm 0.35$). The likelihood that it is the 
correct identification is 
strongly reinforced by a reanalysis of the combined MERLIN+VLA 1.4-GHz  
map of the field which provides a new radio 
detection of HDF850.1 only 0.1\,arcsec 
from the new near-infrared counterpart, and with sufficient positional accuracy
to exclude all previously considered alternative optical candidates.

We have calculated new confidence limits on the estimated redshift of HDF850.1
in the light of the new radio detection, and find $z = 4.1 \pm 0.5$. We have 
also determined the scale-length, and hence estimated the mass of the 
apparently 
nearby (0.5\,arcsec distant) $z \simeq 1$ elliptical galaxy 3-586.0. From this
we calculate that the flux density of HDF850.1 has been boosted by a factor 
of $\simeq 3$ through gravitational lensing 
by this intervening elliptical, consistent with predictions
that a small but significant fraction of blank-field sub-mm sources
are lensed by foreground galaxies. We discuss the wider implications of 
these results for the sub-mm population and cosmic star-formation history.

\end{abstract}


\section{Introduction}
HDF850.1 was the first sub--mm source discovered via unbiased, blank-field
surveys at 850\,${\rm \mu m}$, its 
presence becoming apparent only $\simeq 15$\,hr 
into the $50$ hr SCUBA imaging observation of the Hubble Deep 
Field (HDF) undertaken by Hughes et al. (1998).
It was also the first SCUBA-selected source to be detected in continuum
emission through mm-wavelength interferometry,
with a further $\simeq 40$\,hr of observations, this time with the 
IRAM PdB interferometer,
yielding its position to an accuracy of  $\pm 0.3$\,arcsec 
(Downes et al. 1999).
Frustratingly, however, despite this accurate position and 
the obvious availability of optical data of unparalleled depth and 
resolution, a convincing host galaxy for this bright sub-mm source
has not yet been identified. Indeed, the difficulty experienced in 
finding an optical counterpart for HDF850.1 has even motivated some authors 
to explore the possibility that some of the sub-mm sources uncovered 
via high galactic latitude SCUBA surveys are actually galactic objects
(Lawrence 2001).

It is important to note, however, that even though the search radius 
permitted by the IRAM PdB results is small, this identification failure is 
{\em not} 
due to a lack of potential optical counterparts (see Fig. 1). 
Indeed, both Hughes
et al. (1998) and Downes et al. (1999) noted that the galaxy 3-586.0, one of 
the most obviously red objects in the three-colour image of the HDF, 
is not only 
consistent with the mm/sub-mm position of HDF850.1, but is statistically
rather unlikely to lie so close to the sub-mm/mm source by chance
($p \simeq 0.05$, because 3-586.0 is relatively bright). However, both
Hughes et al. and Downes et al. rejected the possibility that this 
apparently favourable statistical association necessarily implies
that 3-586.0 is the
correct optical identification. This is because, while  
optical--near-infrared photometry of 3-586.0 strongly indicates it 
is a passively evolving elliptical at $z \simeq 1$ (Fern\'{a}ndez-Soto et 
al. 1999; Rowan-Robinson 2001), existing 
sub-mm $\rightarrow$ radio detections and limits yield an estimated 
redshift for the 850\,${\rm \mu m}$ source of $z \simeq 4$, and appear to
exclude $z < 2$ (e.g. Carilli \& Yun 1999, 2000; Dunne et al. 2000).
Furthermore, as argued by Downes et al., it is not expected that such a 
quiescent elliptical should be a strong emitter of sub-mm radiation.
 
Consequently, HDF850.1 has become the classic test case of the extent to which
the technique of estimating redshifts from sub-mm $\rightarrow$radio photometry
can be trusted, and indeed of the whole idea that many/most bright sub-mm
sources lie at high redshift ($z > 2$).

Given the apparent unsuitability of 3-586.0, the two remaining potential 
high-redshift objects in the vicinity of HDF850.1 have both been considered 
as potential identifications. First, Hughes et al. tentatively suggested
3-577.0 as the correct identification because, while statistically quite
likely to lie close to the sub-mm position by chance ($p \simeq 0.3$), it 
appeared to be the only candidate within the SCUBA search radius with 
an estimated redshift in the appropriate range ($z_{\rm est} \simeq 2.9$;
Fern\'{a}ndez-Soto et al. 1999; Rowan-Robinson 2001, and a 
tentative spectroscopic redshift of $z = 3.36$; Zepf et al. 1997). 
Subsequently, the improved IRAM PdB position excluded this object as a possible
counterpart, and Downes et al. concluded in favour of 3-593.1. However, 
the relatively modest estimated redshift of this source 
($z_{\rm est} \simeq 1.75$; 
Fern\'{a}ndez-Soto et al. 1999; Rowan-Robinson 2001) 
means that this option also appears unconvincing.

One might reasonably ask why the correct identification for HDF850.1 
has proved so elusive, given that several other comparably bright 
sub-mm sources have been discovered and unambiguously identified in the 
intervening years (e.g. Smail et al. 1999, Ivison et al. 1998, 2000, 
Gear et al. 2000, Frayer et al. 2000, Lutz et al. 2001). 
The most likely explanation is that HDF850.1 simply lies at a more
extreme redshift than these identified sources.
The strongest hint of this comes from the fact that these now identified 
SCUBA sources have been detected in the radio at $S_{\rm 1.4GHz} > 
50\, {\rm \mu Jy}$, whereas HDF850.1 has evaded radio detection at the level
of $S_{\rm 1.4GHz} < 23\, {\rm \mu Jy}$ (3-$\sigma$).

The radio investigation of HDF850.1 was temporarily confused by
the suggestion of Richards (1999) that the sub-mm source should
be associated with the radio source VLA 3651+1221, 
some 6 arcsec south-west of the 
SCUBA position. This possibility was excluded shortly thereafter 
by the IRAM PdB detection. 
However, this diversion does at least serve to emphasize the 
dangers of making the (initially understandable) assumption that every 
SCUBA source can be reliably associated with
a VLA source provided one is found within a fairly generous search radius
(e.g. Barger et al. 2000), and confirms that any redshift distribution
for the sub-mm population deduced from such SCUBA+VLA associations should be 
regarded as conservative (as noted by Smail et al. 2000).
In fact, to date the most promising radio 
counterpart of HDF850.1 is VLA 3651+1226, 
listed in the supplementary list of sources provided by Richards et al.
(1998) as 
a 2.3-$\sigma$ detection at 8.4\,GHz. Subsequent re-analysis of these data 
has raised the significance of this detection to over 3-$\sigma$, with 
$S_{\rm 8.4GHz} = 7$\,${\rm \mu Jy}$.
However, even if real, this 8.4-GHz detection is still consistent
with the 1.4-GHz limit, and the fact that
the observed radio flux-density from HDF850.1 is lower than that of 
comparably bright sub-mm sources which have already been successfully 
identified.

\begin{figure*}
\vspace*{11.25cm}
\includegraphics{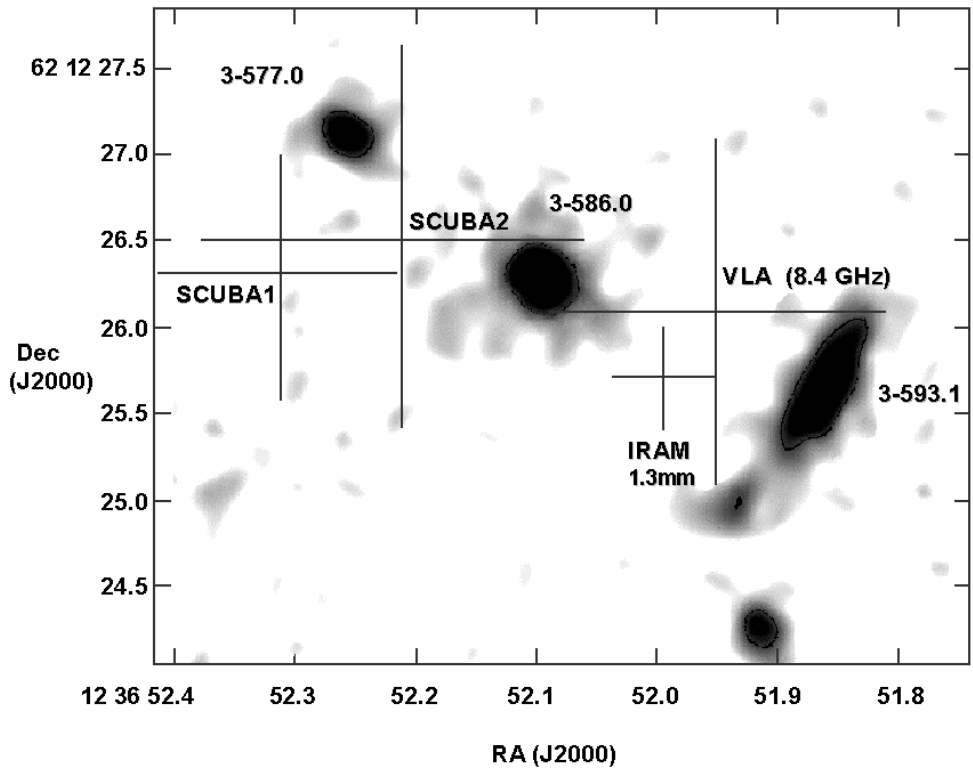}
\caption{\small{Positional information on HDF850.1 available prior to the 
current study. Superimposed on a greyscale of the WFPC2 optical image of 
the relevant region are:
(i) the original position 
of the SCUBA 850\,${\rm \mu m}$ 
source HDF850.1 reported by Hughes et al. (1998) 
(marked {\it SCUBA1}), (ii) the revised position 
of the 850\,${\rm \mu m}$ detection
deduced by Serjeant et al. (2002) (marked {\it SCUBA2}), 
(iii) the position of the tentative
8.4-GHz detection VLA 3651+1226 reported by Richards et al. (1998), and (iv)
the position of the 1.3-mm detection with the IRAM PdB interferometer
reported by Downes et al. (1999). In each case the size of the cross indicates
the 1-$\sigma$ error in the relevant position (see section 2 for further 
details).}} 
\end{figure*}

In this paper we report new results which reveal the host galaxy of 
HDF850.1 to be a faint, extremely red object lying at $z \simeq 4$. 
This conclusion follows from a detailed analysis of a new, deep ($>$10-hr),
high-quality (0.6-arcsec seeing) $K^{\prime}$
image of the HDF obtained with 
the 8.2-m Subaru telescope. This has revealed a new near-infrared counterpart 
to HDF850.1, which we also find to be 
marginally detected in the HST NICMOS F160W 
image of the HDF.
Our conclusion that this new, faint, red object is indeed the host
of the sub-mm source is then strengthened by a re-analysis of the combined
MERLIN+VLA 1.4-GHz radio image of the HDF. This yields a 4-$\sigma$ detection
of HDF850.1 only 0.1 arcsec 
distant from the new near-infrared counterpart, and 
with sufficient positional accuracy
to exclude all the previously-considered alternative optical candidates
discussed above.
We have also utilised this new radio detection in a calculation designed 
to set realistic confidence limits on the estimated redshift of HDF850.1,
and find $z = 4.1 \pm 0.5$. 
Finally, since this new discovery cannot  change the fact that the 
$z \simeq 1$ elliptical 3-586.0 is unlikely to lie so close to the 
sub-mm/mm/radio source by chance, we calculate the probability of the most
likely explanation of this coincidence, namely that HDF850.1 is gravitationally
lensed by 3-586.0. This possibility has been considered before by Hughes 
et al. and Downes
et al. (and indeed a lensing model for 3-586.0 was 
discussed by Hogg et al. (1996) prior to the discovery of HDF850.1). However,
a more accurate calculation can now be performed given the improved 
positional accuracy for HDF850.1, 
new limits on the brightness of a counter image,
and the accurate estimate of the mass of 3-586.0 which we have 
deduced from our detailed modelling of the
infrared and optical images.

The structure of the paper is as follows.
In the next section we briefly 
summarize all the multi-frequency information for
HDF850.1 (and its immediate vicinity) which was available prior to this new 
study. Then, in section 3 we describe the new 
$K^{\prime}$ data, and the results
of the reanalysis of the MERLIN+VLA 1.4--GHz dataset. In section 4 we explain
how we analysed the $K^{\prime}$ image in two independent ways, and demonstrate
that both methods reveal the same new candidate 
identification, which we hereafter refer to as HDF850.1K. We also show  
that a similar analysis of the existing HST NICMOS F160W image of the HDF
yields a marginal $H$-band detection of this object.  

In section 5 we present photometric and astrometric 
information for HDF850.1K, summarize the properties of the elliptical 3-586.0
derived from our image analysis, and present new confidence limits 
on the estimated redshift of HDF850.1. In section 6 we draw this information 
together and provide a quantitative discussion of the likelihood that 
the new near-infrared identification
is indeed the host galaxy of HDF850.1. We also calculate 
the likely extent to which this source 
has been gravitationally lensed by the elliptical 3-586.0. We conclude
with a discussion of the implications of these new results for the sub-mm 
galaxy population in general, and for current estimates of star-formation
density at high redshift.
Throughout this paper we assume a flat cosmology with 
$H_0 = 70 \kms\,{\rm Mpc^{-1}}$, $\Omega_m = 0.3$, 
and $\Omega_{\Lambda} = 0.7$.

\section{Existing Data}

\subsection{Sub--millimetre data}
HDF850.1 was detected by Hughes et al. (1998) at 850\,${\rm \mu m}$ 
with a flux density of $S_{\rm 850\mu m} = 7.0 \pm 0.5$\,mJy. 
This flux density, 
derived from the map, was also confirmed to within the errors by 
pointed photometric observations. The position of this 
object as derived from the original SCUBA map of the HDF was (J2000)\\

\noindent
RA 12$^h$ 36$^m$ 52.32$^s$\ \ Dec +62$^\circ$ 12$^{\prime}$ 26.3$^{\prime \prime}$\\

\noindent
The formal uncertainty in this position, based on the SCUBA beam size at 
850\,${\rm \mu m}$, and the S/N of the detection was 
quoted by Hughes et al. to be 0.7\,arcsec (1-$\sigma$) in each 
dimension. This position, and the 1-$\sigma$ 
errors,
are marked by a cross in Fig. 1, and labelled `SCUBA1'. 

A reanalysis of the HDF SCUBA map has recently been completed by Serjeant 
et al. (2002), using the maximum-likelihood source extraction 
technique described by Scott et al. (2002). This yields a slightly different
position for HDF850.1 (albeit one consistent with the originally 
quoted formal uncertainty) which is (J2000):\\

\noindent
RA 12$^h$ 36$^m$ 52.22$^s$\ \ Dec +62$^\circ$ 12$^{\prime}$ 26.5$^{\prime \prime}$\\

\noindent
Serjeant et al. (2002) also performed a series of simulations to estimate
the effect of confusion noise on extracted source position in the SCUBA HDF 
map, and found that, even at the 7\,mJy level, the mean positional offset 
is $\simeq 1.5$\,arcsec. 
In Fig. 1 we therefore also show this revised SCUBA position for HDF850.1,
labelled `SCUBA2', with a positional error of 1.5\,arcsec (1.1\,arcsec in 
each dimension).
 
Hughes et al. also derived a 450\,${\rm \mu m}$ 3--$\sigma$ 
upper limit of $S_{\rm 450\mu m} < 21$\,mJy. As discussed by 
Hughes et al., a non-detection at this level 
constrains the redshift of HDF850.1 to $z > 2.5$, for the known range of 
far-infrared--to--sub-mm spectral energy distributions.  
 
Finally, we note that Serjeant et al. quote a somewhat smaller value 
for the 850\,${\rm \mu m}$ flux density of HDF850.1 than Hughes et al. (i.e.
$S_{\rm 850\mu m} = 5.6 \pm 0.4$\,mJy). However, in this paper we adopt 
the originally quoted value of 7\,mJy, because this value is more
consistent with the result derived from the pointed photometric observations.
 
\subsection{Millimetre data}
HDF850.1 was also detected by Hughes et al. at 1.35\,mm, and found to 
have a flux density of $S_{\rm 1.35mm} = 2.1 \pm 0.5$\,mJy. This observation 
was made with a beam of FWHM 23\,arcsec, and so can safely be regarded 
as a measure of the total 1.35\,mm flux density of the source.
Given this, the subsequent measurement of $S_{\rm 1.3mm} = 2.2 \pm 0.3$\,mJy
by Downes et al. (1999) within the $\simeq 2$-arcsec beam of the IRAM PdB 
interferometer, indicates that the source is reasonably 
compact.

The 1.3-mm position measured for HDF850.1 by Downes et al. is (J2000)\\

\noindent
RA 12$^h$ 36$^m$ 51.98$^s$\ \ Dec +62$^\circ$ 12$^{\prime}$ 25.7$^{\prime \prime}$\\

\noindent 
with an uncertainty of 0.3 arcsec (1-$\sigma$) in each dimension
This position, and the 1-$\sigma$ errors,
are also marked by a cross in Fig. 1. This position is clearly   
consistent with the SCUBA position in the light of the revised uncertainty 
in the SCUBA astrometry discussed above. It is 1.9\,arcsec distant 
from the SCUBA position of HDF850.1 derived by Serjeant et al. (2002), 
corresponding to the 69th percentile of the distributions of 
positional offsets estimated from the simulations of the effects
of confusion noise on source position.

\subsection{Radio data}
The formal position of the only possible marginal radio 
detection of HDF850.1 (i.e. the 
$\simeq$\,3-$\sigma$ $S_{\rm 8.4GHz} \simeq 7$\,${\rm \mu Jy}$ 
`source' VLA 3651+1226 from the supplementary 
list of sources in Richards et al. 1998) is (J2000)\\ 

\noindent
RA 12$^h$ 36$^m$ 51.96$^s$\ \ Dec +62$^\circ$ 12$^{\prime}$ 26.1$^{\prime \prime}$\\

\noindent
but the low significance of this detection means that the 1-$\sigma$ error in
this position is $\simeq 1$\,arcsec in each dimension.
This position and the 1-$\sigma$ errors
are marked by the large cross in Fig. 1. It can be seen that the position of
this radio detection is consistent with both the SCUBA and IRAM positions.

\subsection{Optical data}
Finally, also shown (and labelled) in Fig. 1 are the 3 optical objects 
3-577.0, 3-586.0, and 3-593.1 which, as discussed in the introduction, have 
been previously considered as potential identifications by various authors.
Unfortunately, a spectroscopic redshift has yet to be 
determined for any of these objects despite, at least in the case of 
3-586.0, deep optical spectroscopy with the 10-m Keck telescope
(Stern, private communication). The colour-estimated redshifts for these
3 objects are $z_{\rm est} \simeq 2.9 $, $z_{\rm est} \simeq 1.1 \pm 0.1$ and 
$z_{\rm est} \simeq 1.7$ respectively 
(Fernandez-Soto et al. 1999; Rowan-Robinson 2001).
In the case of 3-586.0 the estimated redshift is relatively secure due to
the fact that its broad--band SED so clearly mimics that expected from an evolved
elliptical at $z \simeq 1$, with a clear break between the $I$ and $J$ bands. 
The apparent passivity of this elliptical is also the most likely 
explanation for the lack of any discernable emission-line features in its
optical spectrum. 

The positions of the optically-detected galaxies shown in Fig. 1 are as
given in Williams et al. (1996) and Downes et al. (1999). However, the 
$\pm 0.4$-arcsec errors on RA and Dec given in these papers can now be revised
down to $\pm 0.1$\,arcsec following the registration of the HDF 
optical image with the MERLIN image (see section 3.2) to within an accuracy 
of 50\,milliarcsec.
 
\section{New observations}

\subsection{Deep Subaru ${\bf K^{\prime}}$ Imaging}

The new deep $K^{\prime}$ image of the HDF-N was obtained with the Subaru
8.2-m telescope equipped with the NIR camera CISCO (Motohara et al. 1998) on
April 3--4, 2001. Full details of the observations and data reduction will be 
given in a future publication (Kajisawa et al., in prep.). Here we 
briefly summarize the data properties. The net on-source integration time is 
10.2\,hr. The data were reduced in a standard manner; each individual 
frame was flat-fielded and median-sky subtracted before the frames were 
combined. Photometric calibration was peformed 
using UKIRT Faint Standard Stars observed at 
various altitudes during the observation, to produce $K$ magnitudes
in the UKIRT Mauna Kea system (Hawarden et al. 2001).
We have ignored the $K^{\prime} \rightarrow K$ colour correction, which 
introduces an 
uncertainty in the $K$ magnitude of the detected sources of 
at most 0.1\,mag depending on their spectra.  The original pixel
scale of CISCO is 0.111 arcsec, but the images have been carefully 
resampled to match
the WFPC2 drizzeled data (see subsection 4.1 below). The seeing was 
$0.4-0.6$\,arcsec during the observations and the FWHM of stellar objects in 
the final image is $\simeq 0.6$\,arcsec. The peak of the number counts
of the detected sources reaches a magnitude of $K \simeq 23$. A greyscale
representation of the $8 \times 8$\,arcsec sub-region of this
image, centred on 3-586.0, is shown in the top panels of Fig. 2.

\subsection{MERLIN+VLA 1.4-GHz image}
            
18\,days of MERLIN data and 42\,hours of A--array VLA data at 
1.4\,GHz have been 
combined to image radio sources in a 10-arcmin field centred on the HDF. 
This area includes both the HDF and the Hubble 
Flanking Fields (HFF).  These are the most sensitive 1.4-GHz images yet made, 
with rms noise levels of 3.3\,${\rm \mu Jy}$/beam 
in the 0.2-arcsec  resolution images 
(Muxlow et al. 2002).  Positions derived from the independent MERLIN and 
VLA imaging are with respect to the International Celestial Reference Frame 
(ICRF, Ma et al. 1998), and agree to better than 15\,milliarcsec over the 
entire 10\,arcmin field. Radio sources associated with compact galaxies 
have been used to align the HST WFPC2 optical fields to better than 
50\,milliarcsec in the HDF itself, and to better than 
$\simeq 150$\,milliarcsec in the outer parts of the 
HFF (Muxlow et al. 
2000).  This astrometric alignment of the HST WFPC2 fields has been used by 
Garrington, Muxlow \& Garrett (2000) to argue that the position of 
HDF850.1 as measured by the IRAM telescope (Downes et al. 1999), again 
with respect to the ICRF, precludes its identification with any galaxy 
visible on the HDF WFPC2 frame.

In that region of the HDF close to the position of HDF850.1, the 
high-resolution MERLIN+VLA image has been smoothed to 0.6-arcsec resolution. 
In the smoothed image, an area of extended radio emission is detected 
between the elliptical galaxy 3-586.0 and the IRAM position for HDF850.1.  
This detection is shown by the contours in Fig. 4a, where it can be seen
that the  radio emission splits roughly into two components, one partially 
overlaying the central region of 3-586.0, and the other lying 0.5\,arcsec
to the East of the IRAM position.  Fitting a two-dimensional Gaussian to the 
latter radio component yields a flux of 16$\,{\rm \mu Jy}$ at the 
position given in Table 2.  With an rms noise level of 3.7\,${\rm \mu Jy}$/beam
in the smoothed radio image, the source detection is significant at  
better than the 4-$\sigma$ level.

\section{Near-Infrared Image Analysis}

The raw Subaru $K^{\prime}$ image of the field 
shown in the top panel of Fig. 2 contains 
no obvious new candidate identification for HDF850.1 over and above 
those present in the HST WFPC2 optical images as discussed in Section 2.
However, because of the extreme depth of this image, and the
fact it has been taken from the ground (albeit in 
excellent conditions, with a seeing disc FWHM of 0.6\,arcsec) 
the objects in the vicinity of 
HDF850.1 appear considerably 
more extended than in the corresponding HST images. 
In particular, near-infrared emission from the 
$z \simeq 1$ elliptical 3-586.0 is detectable out to an angular radius of 
$\simeq 1.5$\,arcsec, and thus covers a significant region of the image
within which the true identification of HDF850.1 could potentially lie.
We therefore decided to attempt to remove the light of 3-586.0 from 
the $K^{\prime}$ image. This was done in two independent ways, first by 
subtracting an appropriately blurred and scaled version of the F814W 
image of the field, and second by fitting and removing a model representation
of the elliptical from the $K^{\prime}$ 
image. As described below, these two 
alternative approaches yielded re-assuringly similar results.
We have also performed a similar analysis on the HST NICMOS Camera-3 F160W 
image of the HDF (Dickinson et al. 2002) and find that this yields 
a marginal $H$-band detection at the same position
(to within $< 0.1$\,arcsec) 
as the $K^{\prime}$ detection.
   
\subsection{Subtraction of the F814W image from the ${\bf K^{\prime}}$ image}

The $K^{\prime}$ image was interpolated onto the same pixel scale as the 
drizzled F814W image (0.04\,arcsec), and then 
$8 \times 8$\,arcsec postage-stamp
images were extracted at both wavelengths, centred on the fitted centroid
of the elliptical 3-586.0. The F814W image was then convolved with a Gaussian, 
scaled to the same central peak height as the $K^{\prime}$ image, and  
subtracted from it. The FWHM of the convolving Gaussian was varied until 
optimally clean subtraction of 3-586.0 from the residual image was 
achieved. The 
optimum FWHM was 0.57\,arcsec which, when added in quadrature to the 
intrinsic width of the HST F814W point spread function, makes good sense
given the rms seeing of 0.6\,arcsec determined from the full 
$K^{\prime}$ image
of the HDF.

This process is summarized in Fig. 2 (left-hand column) 
which shows the $K^{\prime}$ image, the 
convolved F814W image, and the residual image. The small cross in the 
residual image marks the position of the centroid of 3-586.0. It can be seen
that the brightest source in this pseudo $K-I$ residual flux-density image 
is a small source, of visible angular extent $\simeq 0.5 - 0.7$\,arcsec, lying 
0.55\,arcsec distant from the projected centre of 3-586.0.

\begin{figure*}
\vspace*{19.7cm}
\includegraphics{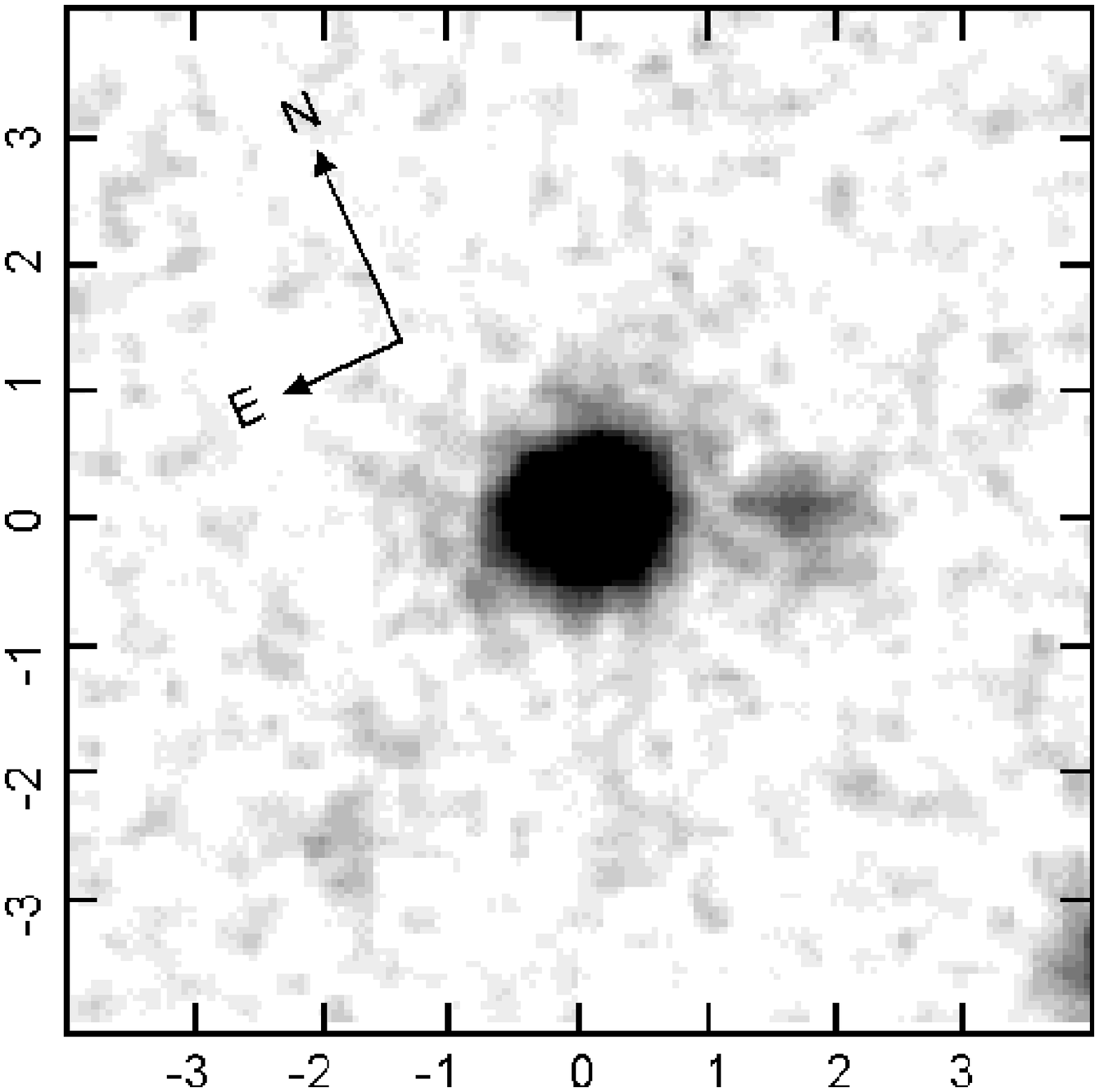}
\includegraphics{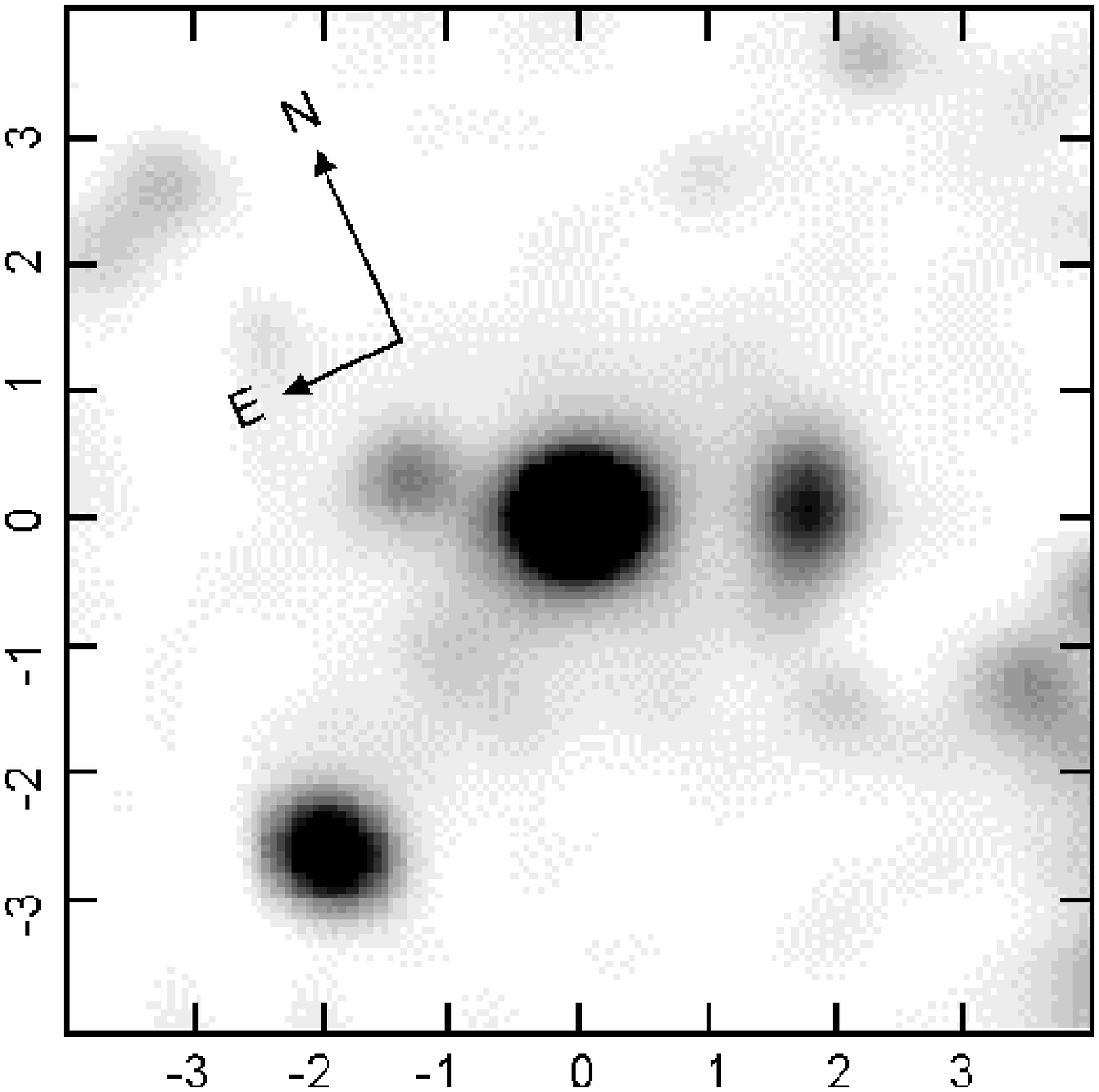}
\includegraphics{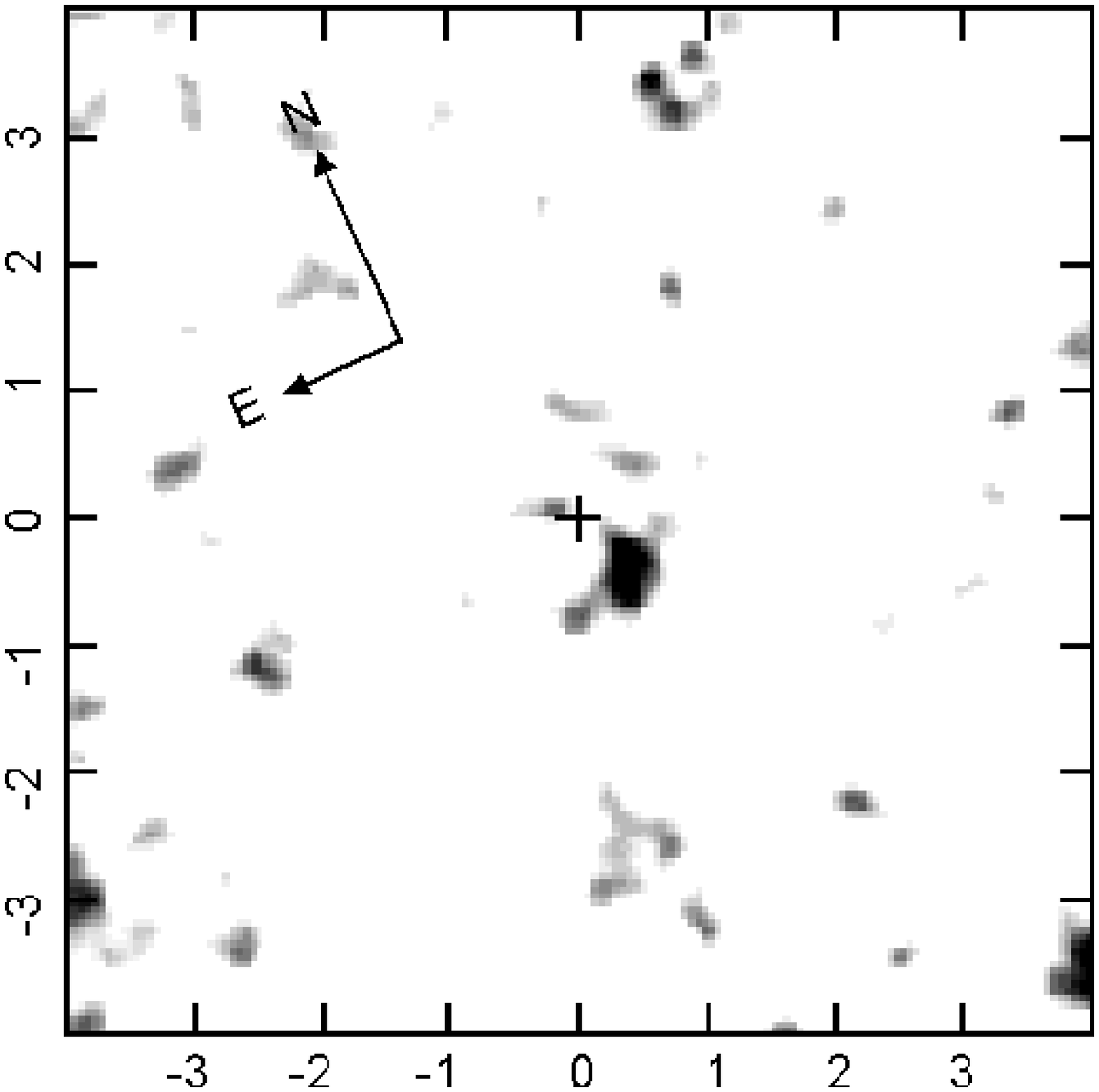}
\includegraphics{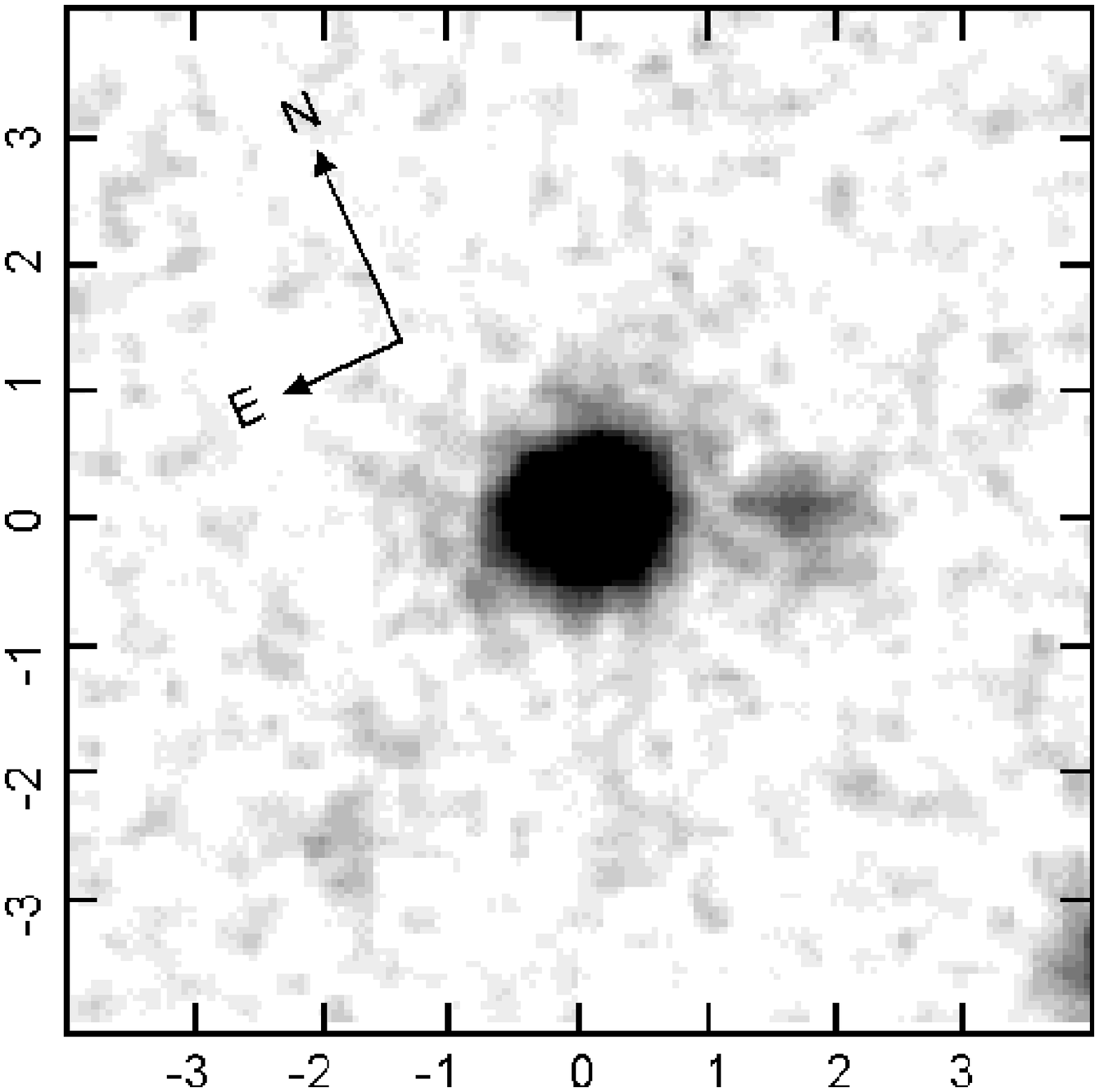}
\includegraphics{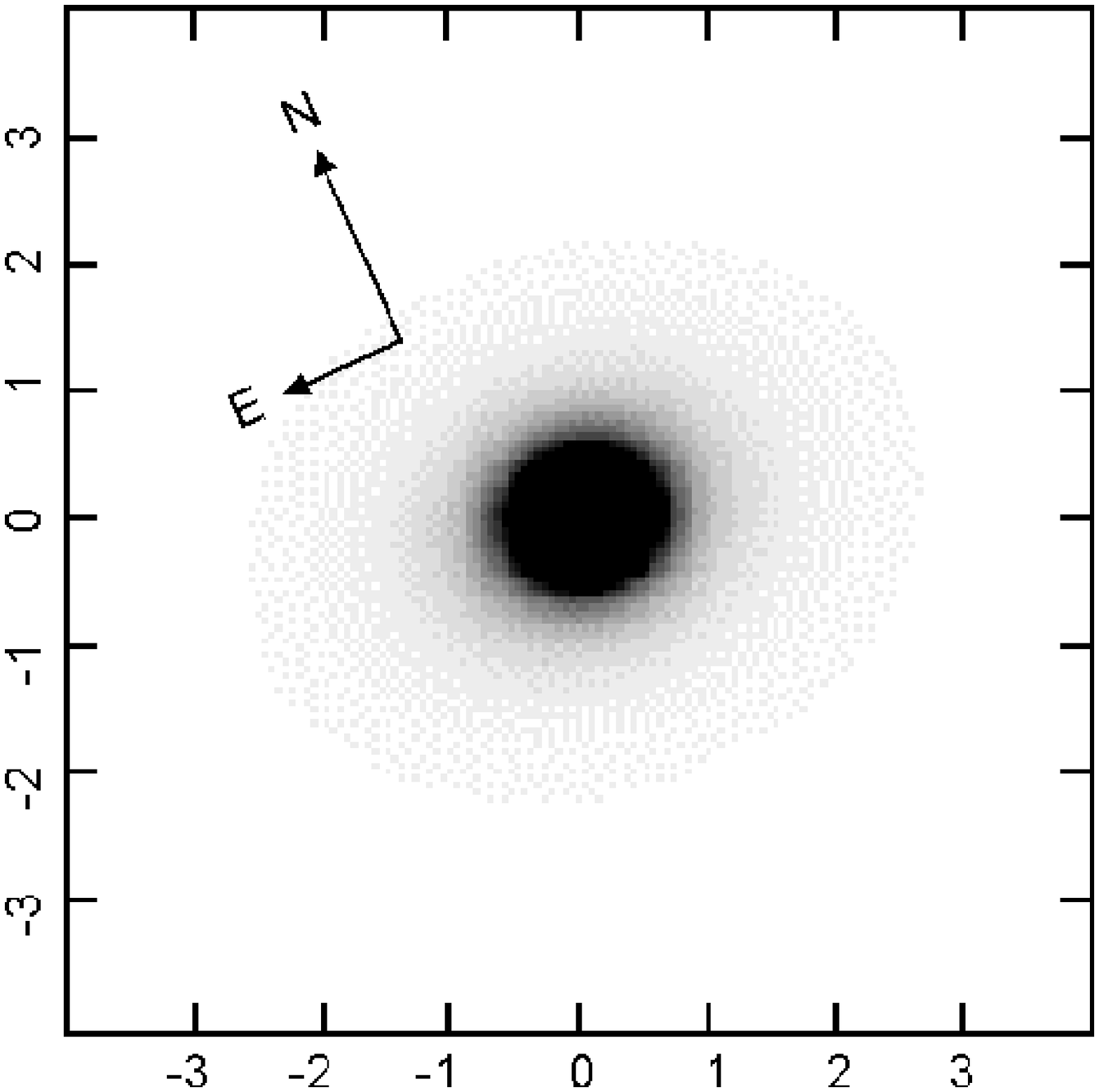}
\includegraphics{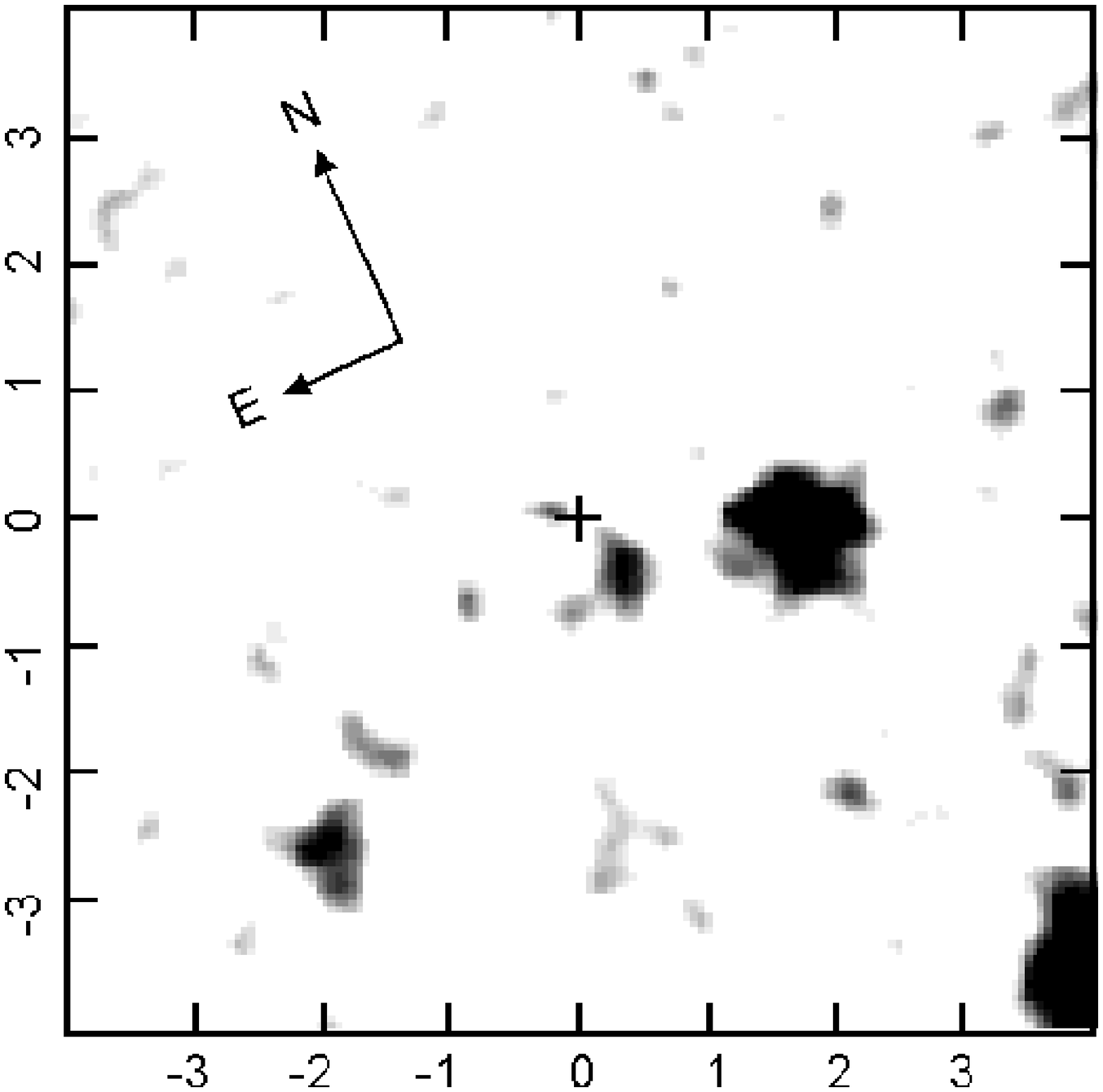}
\caption{\small Subtraction of the $z \simeq 1$ elliptical 3-586.0 from
the Subaru $K^{\prime}$ image. The left-hand column shows 
the result of doing 
this using the HST WFPC2 F814W image of the same field. The 
top-left panel shows an $8 \times 8$\,arcsec postage stamp extracted from the 
Subaru image, centred on 3-586.0. The centre-left 
panel shows the same 
region, as imaged by the HST in the F814W filter, after convolution 
with a Gaussian of FWHM 0.57\,arcsec. The bottom-left panel shows the residual 
image which results from subtracting the middle (F814W) image from the 
top ($K^{\prime}$) image after scaling 
both (background-subtracted) images to
the same peak pixel value. 
The residual image has been smoothed with a Gaussian of FWHM 0.2\, arcsec, 
and the cross marks the position of the centroid 
of 3-586.0 prior to subtraction.
The right-hand column shows the result of doing
this by fitting and subtracting the best-fitting de Vaucouleurs model
of the $K^{\prime}$ light from 3-586.0. The top-right panel again shows 
the $K^{\prime}$ postage stamp. The centre-right panel shows the 
best-fitting 
elliptical galaxy model for 3-586.0, after convolution with 0.6-arcsec
seeing (this model has an axial ratio of 1.28, a position angle 
of 85 degrees east of north, and a half-light radius of 3.0\,kpc). The 
bottom-right panel shows the residual image which results from subtracting
the middle (model) image from the top (observed $K^{\prime}$) image.
Again the residual image has been smoothed with a Gaussian of FWHM 0.2\,arcsec,
and the cross marks the centroid 
of 3-586.0 prior to subtraction.}
\end{figure*}

\begin{figure}
\vspace*{19.7cm}
\includegraphics{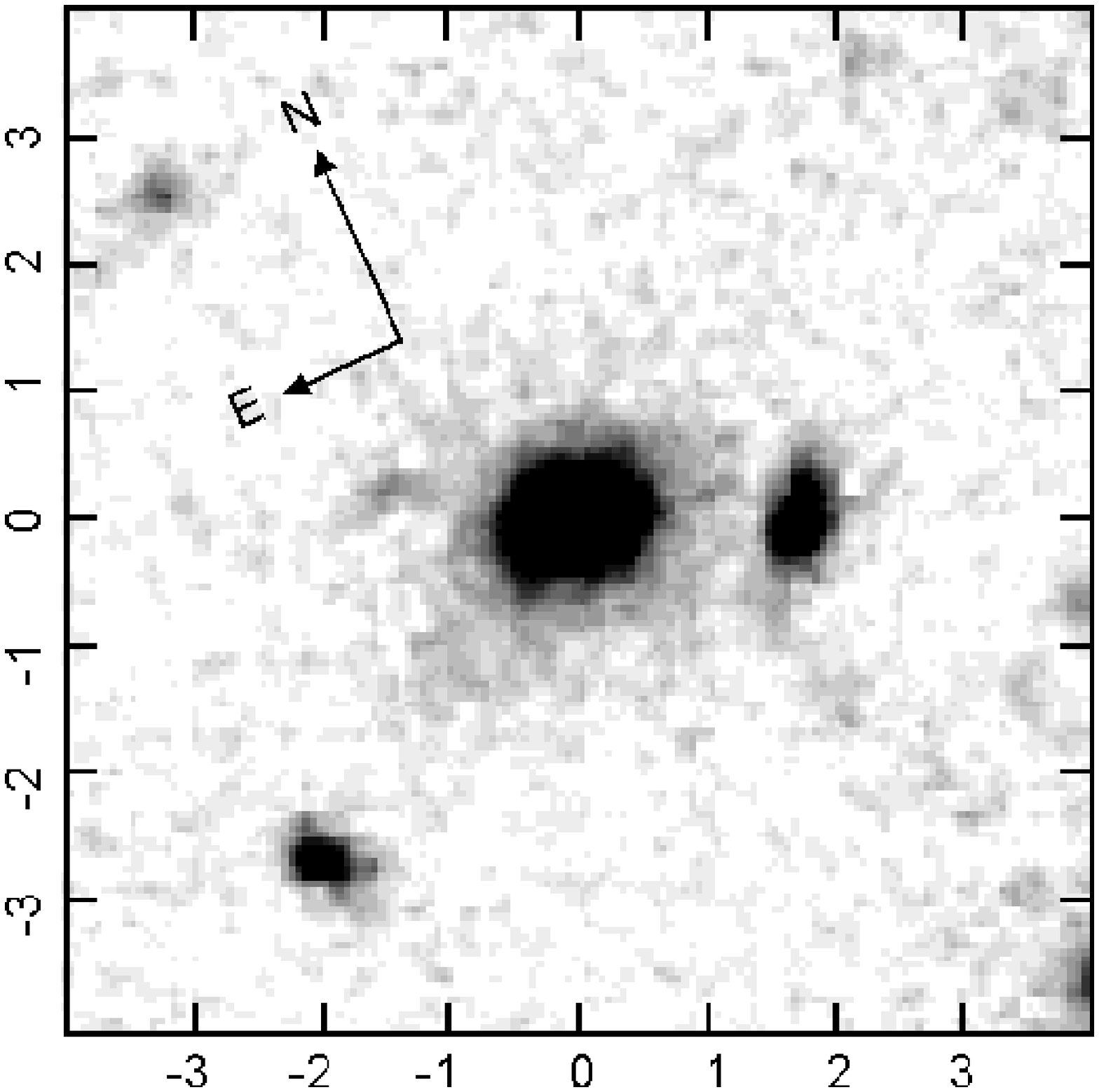}
\includegraphics{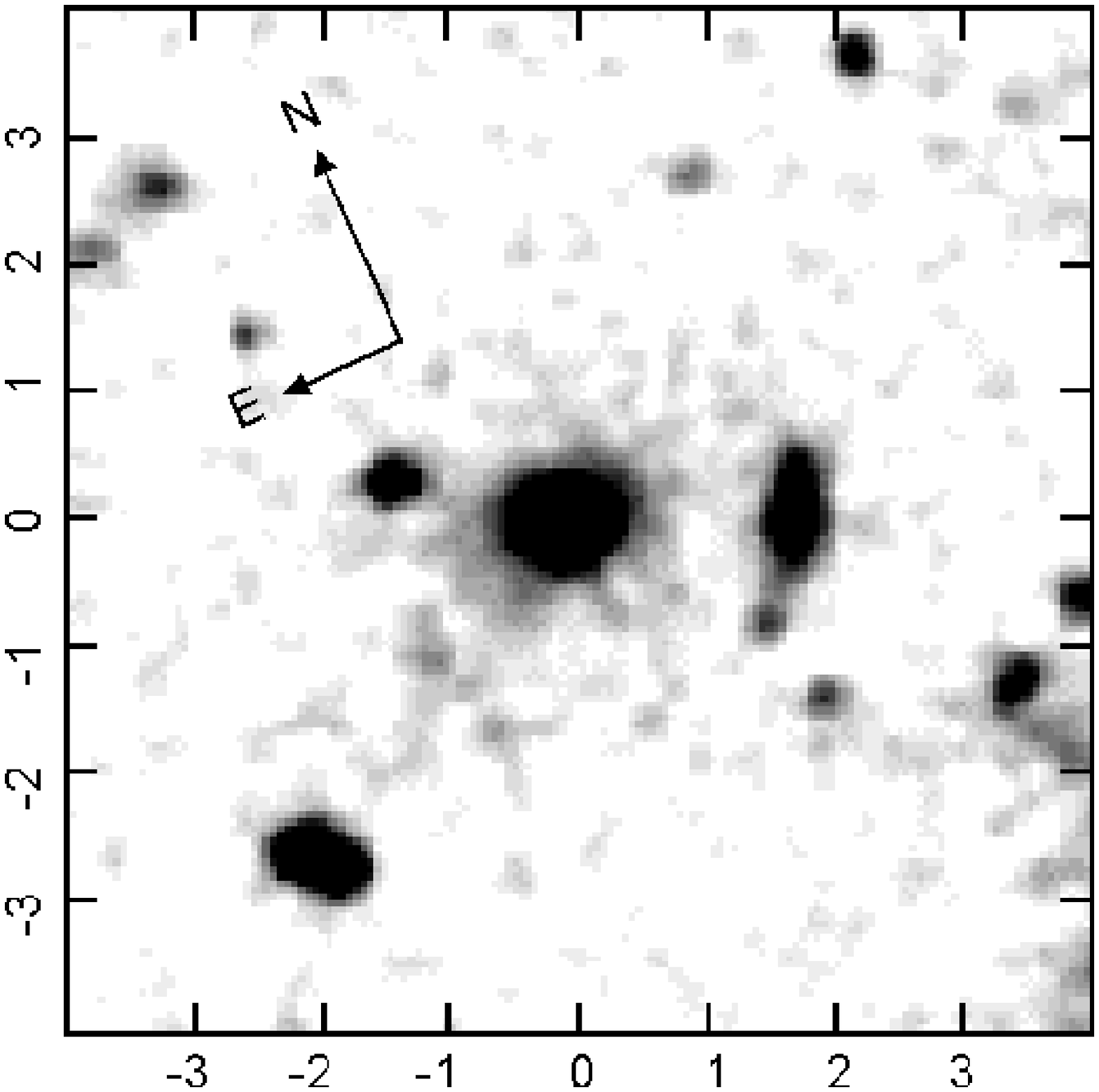}
\includegraphics{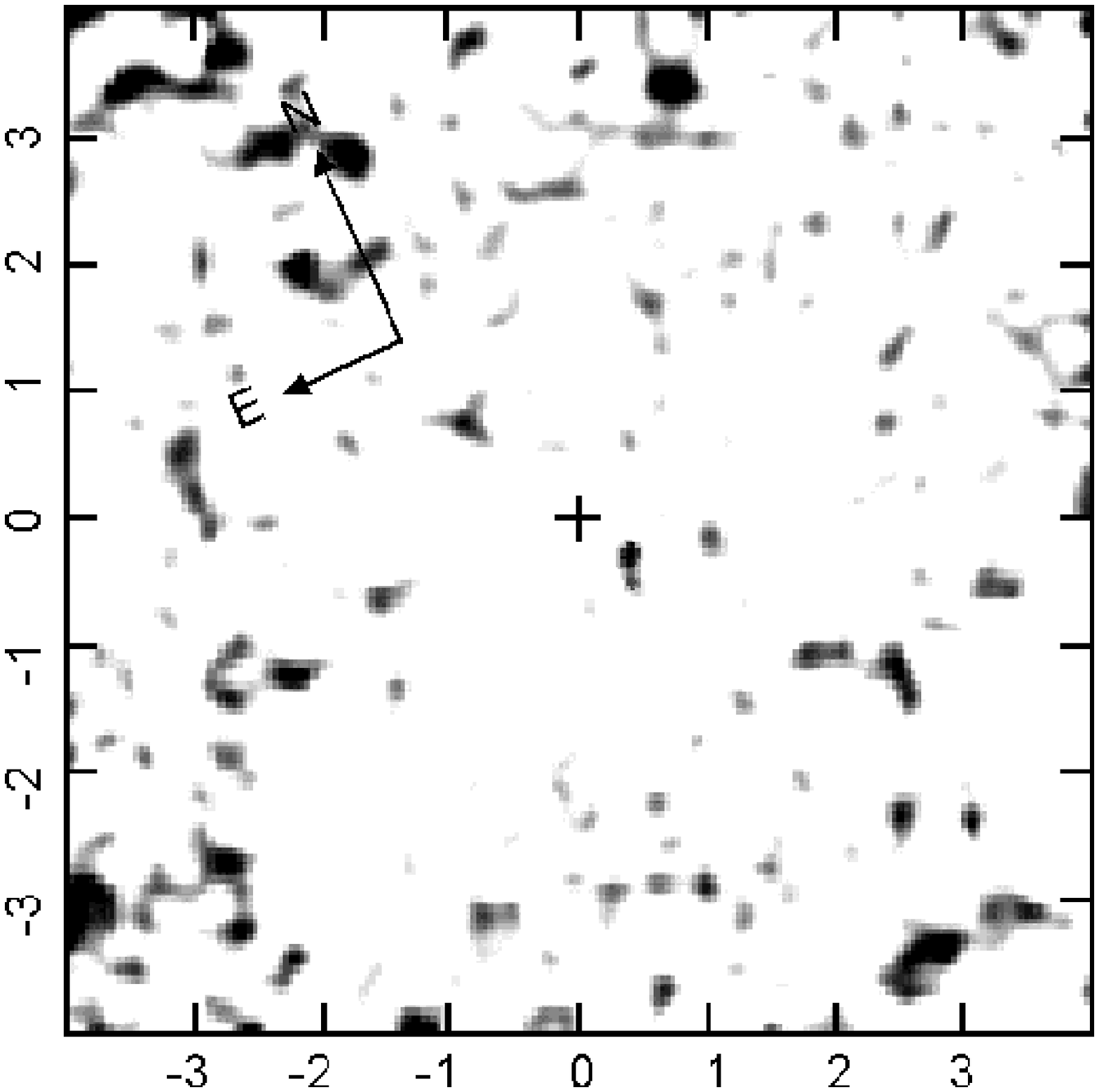}
\caption{\small Subtraction of the 
$z \simeq 1$ elliptical 3-586.0 from
the HST NICMOS F160W image using a smoothed version of 
the HST WFPC2 F814W image of the same field. The 
top panel shows an $8 \times 8$\,arcsec postage stamp extracted from the 
NICMOS image, centred on 3-586.0. The data have been ``drizzled'' 
onto a pixel scale of 0.04\,arcsec, for ease of comparison with the 
WFPC2 data. The centre panel shows the corresponding 
F814W image, after convolution 
with a Gaussian of FWHM 0.19\,arcsec. The bottom panel shows the residual 
image which results from subtracting the middle (F814W) image from the 
top (F160W) image after scaling 
both images to
the same peak pixel value. The residual image has been smoothed 
with a Gaussian of FWHM 0.12\,arcsec, and the 
cross marks the centroid 
of 3-586.0.}
\end{figure}

\subsection{Subtraction of a  model elliptical from the ${\bf K^{\prime}}$ image}
A model elliptical galaxy was fitted to the original 0.1-arcsec pixel 
Subaru image of 3-586.0 using a 2-dimensional modelling code 
originally developed for modelling quasar host galaxies (McLure, Dunlop
\& Kukula 2000), incorporating 0.6-arcsec seeing. This model elliptical was 
then simply subtracted from the image to produce a second, alternative 
residual map of the field which was then interpolated onto 0.04-arcsec 
pixels for ease of comparison with the pseudo $K-I$ image discussed in the
previous subsection.

This process is also summarized in Fig. 2 (right--hand column), 
which shows the raw $K^{\prime}$ image,
the image of the best fitting de Vaucouleurs model of 3-586.0, 
and the residual
image. All objects detected in the $K^{\prime}$ image 
other than 3-586.0 are 
obviously still present in this residual image, but  
the one new object revealed by this model subtraction process is once again
a small source, of visible angular extent $\simeq 0.5 - 0.7$\,arcsec, lying 
0.55\,arcsec distant from the projected centre of 3-586.0.

\subsection{Subtraction of the F814W image from the NICMOS F160W image}

The drizzled F160W image of the HDF (Dickinson et al. 2002)
was interpolated onto the same pixel scale as the 
drizzled F814W image (0.04\,arcsec), and then 
$8 \times 8$\,arcsec postage-stamp
images were extracted at both wavelengths, centred on the fitted centroid
of the elliptical 3-586.0. The F814W image was then convolved with a Gaussian
of FWHM 3.5 pixels (in an attempt to degrade it to a resolution comparable
to that of the F160W image), 
scaled to the same central peak height as the F160W image, and  
subtracted from it. The result of this process was not as clean as that 
achieved using the $K^{\prime}$ image. In particular it  
produced a region of oversubtraction
around the centroid of 3-586.0 (at a radius $\simeq 0.1 - 0.2$\, arcsec)
presumably due to the difference 
between the Gaussian-convolved WFPC2 PSF and the more Airy disc-like 
NICMOS PSF. However, attempts to resolve this problem using convolution 
with the F160W PSF predicted by the {\sc tiny tim} software did not 
produce any significant improvement. The resulting residual image should 
therefore be regarded with some caution. Nevertheless, as shown in Fig. 3,
(which shows the NICMOS F160W image, the 
convolved F814W image, and the residual image) the brightest source 
in the central region of this pseudo $H-I$ 
residual flux-density image lies in exactly the same position, and has the 
same basic shape as the source revealed by the analysis of the $K^{\prime}$
image discussed above and illustrated in Fig. 2.

\section{Results}

\subsection{Discovery of HDF850.1K, the faint ERO host of 
the sub-mm source HDF850.1}

The detection of a new source 0.55\,arcsec distant from 
3-586.0 in the bottom-right panel of Fig. 2 shows that it is
genuinely due to an excess of $K$-band flux-density, rather than some defect
in the F814W image. Conversely its presence in the bottom-left panel 
of Fig. 2
demonstrates that it is not simply an artefact of the model-fitting process
due to, for example, the galaxy 3-586.0  not being well described by a smooth 
elliptical model. Moreover, the similarity of this object in these two 
alternative residual images ($K-K_{\rm model}$ and $K-I$) means that 
this object must be undetected in the F814W image. 

To quantify the significance of this new $K$ detection we have performed
aperture photometry on both the alternative residual images described above.
To quantify the significance of its non-detection in the $I$-band, we have 
also performed aperture photometry at the same position in a residual
F814W image, produced by fitting and subtracting 
a 2-dimensional elliptical model to the F814W image of 3-586.0. To obtain 
an estimate of its $H-K$ colour we have performed aperture photometry on 
the residual NICMOS F160W image shown in the bottom panel of Fig. 3,  
centred on the position of the $K^{\prime}$ detection.

The results of this photometry are given in Table 1. In summary, this new 
object (which we hereafter call HDF850.1K) is a faint, but 
clearly significant 
detection at $K$ with $K = 23.5 \pm 0.2$, and is an extremely red 
object, with $I-K > 5.2$ and $H-K = 1.4 \pm 0.35$. 

As can be seen from Table 1, the faintness of HDF850.1K
means that its $K$ magnitude has to be based on measurements made 
through relatively small software apertures ($0.5-1.0$ arcsec). There is no
evidence that HDF850.1K is significantly more extended than the 0.6-arcsec
seeing disc in the sense that, correcting for the effects of seeing, the 
0.5-arcsec measurement is certainly consistent with the 1.0-arcsec aperture
measurement. However we cannot rule out the possibility that, given 
imaging of sufficient depth, HDF850.1K might prove to be as extended 
as the $K$-band 
identification of Lockman850.1 recently reported by Lutz et al.
(2001) (for which the 3-arcsec 
diameter aperture magnitude is 1.6\,mag. brighter
than the 1-arcsec diameter aperture magnitude).
 
After clockwise rotation of the images shown in Fig. 2 to align
with the RA and Dec co-ordinate system, a colour version of the 
residual $K-I$ image has been combined with the information 
presented in Fig. 1, along with the MERLIN+VLA contours discussed in 
section 3.2, to produce
the combined image shown in Fig. 4a. The accurate alignment of the
residual colour image with Fig. 1 is aided by the negative images 
of 3-557.0 and 3-593.1 present in the former. 

\begin{figure*}
\vspace*{21.5cm}
\includegraphics{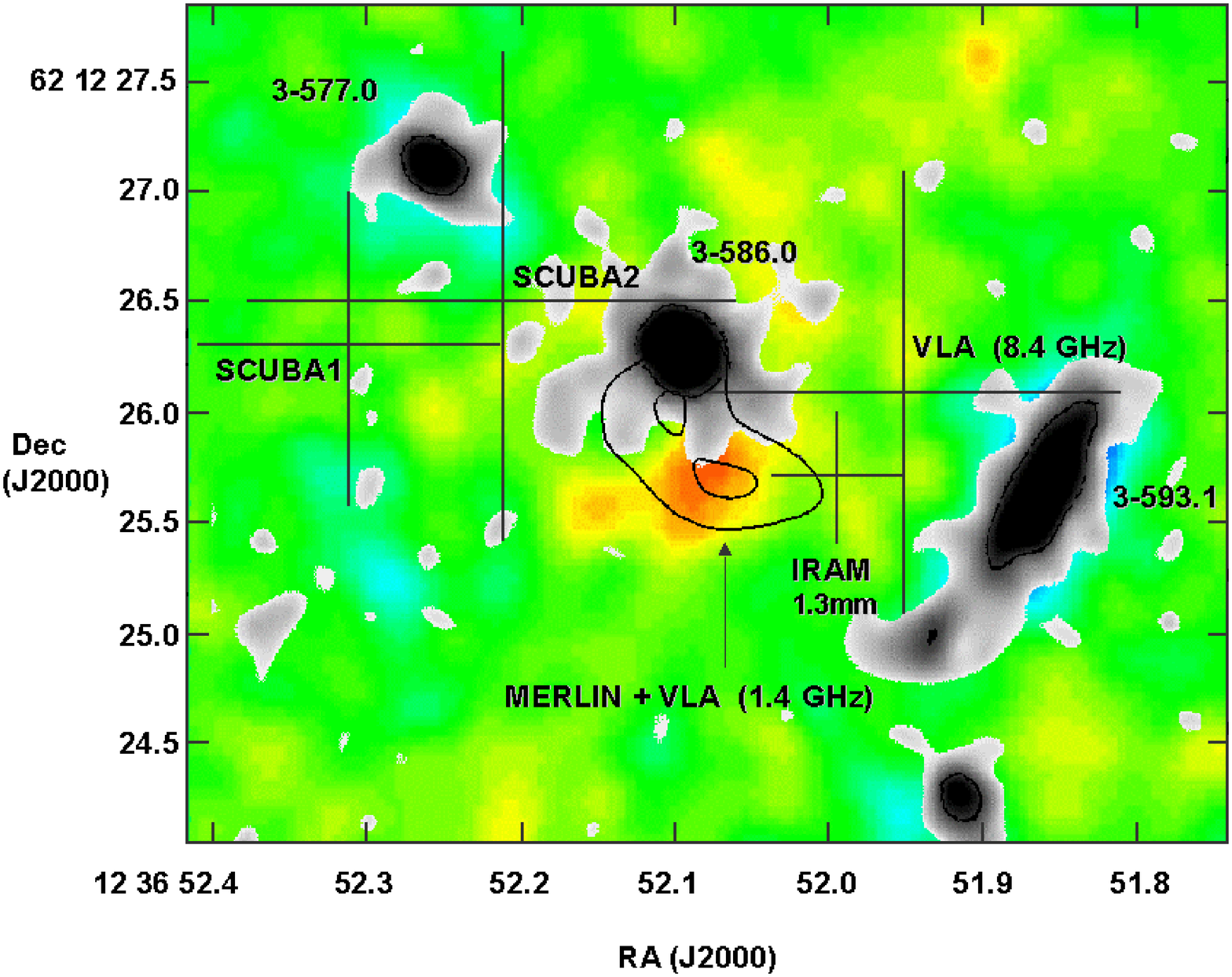}
\includegraphics{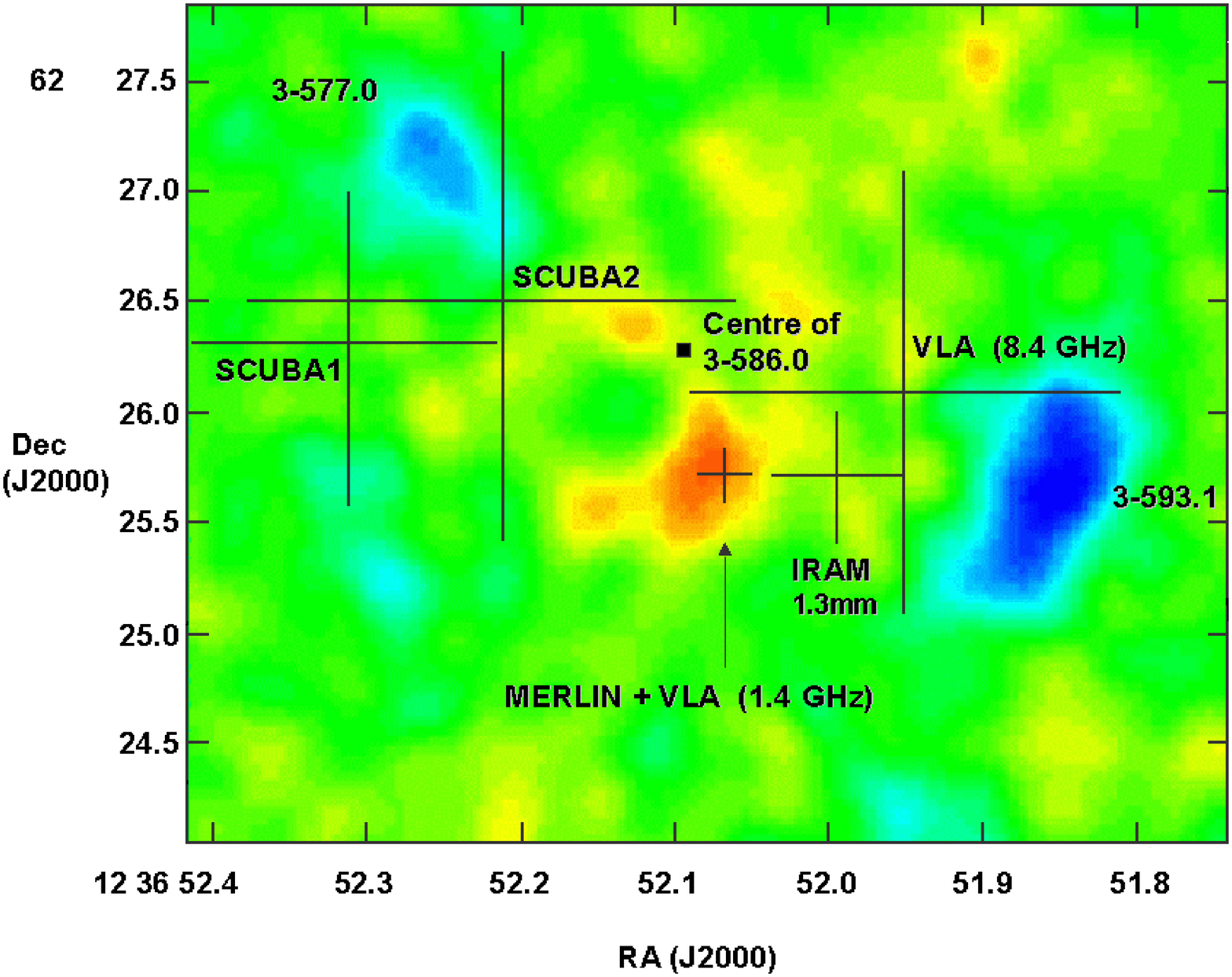}
\caption{\small{Upper Panel: The pre-existing optical-to-radio information
on HDF850.1 from Fig. 1 (with the new MERLIN+VLA 1.4-GHz detection 
illustrated by the contours plotted at 8.5\,${\rm \mu}$Jy/beam, 
$1.5 \times 8.5$\,${\rm \mu}$Jy/beam, $2 \times 8.5$\,${\rm \mu}$Jy/beam etc)
overlaid on the $K-I$ colour image of the field discussed in section 
5.1.
Lower Panel: The same image, but this time with the greyscale representation
of the HST WFPC2 image removed, and the MERLIN+VLA contours replaced by 
a cross marking the formal position of (and 1-$\sigma$ uncertainty in)
the brightest component of the 1.4-GHz source. This image makes it clear
that not only is the position of HDF850.1K consistent with that of the
MERLIN+VLA detection, but that given the $\simeq 0.1$\,arcsec uncertainty 
in the radio position, HDF850.1K is the only plausible identification in 
the field.}}
\end{figure*}

Relative to the centre of 3-586.0, the centroid of HDF850.1K lies 
$0.2 \pm 0.08$\,arcsec west, and $0.5 \pm 0.08$\,arcsec south of 
the elliptical galaxy. Use of the 
accurate (50\,milliarcsec rms) reference frame established from matching the 
MERLIN and HST images of the HDF then yields a position for HDF850.1K of 
(J2000)\\

\noindent
RA 12$^h$ 36$^m$ 52.072$^s \pm 0.015^s$\ \ 
Dec +62$^\circ$ 12$^{\prime}$ 25.75$^{\prime \prime} \pm 
0.1^{\prime \prime}$.\\

\noindent
Given the IRAM PdB position for HDF850.1, and the well-documented fact
(see Hughes et al. 1998 and Downes et al. 1999) that none of the other
plausible optical/IR counterparts (3-586.0, 3-593.1 or 3-593.2) appear
to have estimated redshifts consistent with that estimated for HDF850.1,
it seems highly probable that HDF850.1K is indeed the long-sought host galaxy
of the brightest SCUBA source in the HDF. This probability is obviously
further increased by the fact this object is an ERO, with a colour
very similar to that which has been found for other, sucessful SCUBA
source identifications (e.g. Lutz et al. 2001; Ivison et al. 2000, 2001;
Smail et al. 1999; Frayer et al. 2000).
However, perhaps most impressive of all, as illustrated 
in Fig. 4b, is the essentially exact 
astrometric coincidence of HDF850.1K with the tentative detection of HDF850.1
in the combined MERLIN+VLA 1.4-GHz image discussed above in section 3.2
which yields a new radio position for HDF850.1 accurate to 0.1\,arcsec. 
The astrometric information for the various detections of
HDF850.1 as a function of frequency is summarized in Table 2.

The probability that HDF850.1K is the correct identification of HDF850.1 is 
quantified and discussed in more detail in Section 6.1, while the possible 
effects of lensing by 3-586.0 are discussed in Section 6.2. However, to
better inform this discussion we first consider what can be learned about
3-586.0 from the optical-infrared image analysis presented here, and 
also calculate what new constraints can be placed on the estimated redshift 
of HDF850.1 incorporating its new detection at 1.4\,GHz.
 
\begin{table}
\caption{\small Near-infrared--to--radio photometry of HDF850.1}
\begin{tabular}{|l|l|l|} \hline \hline
 Band & Flux/Magnitude& Method/Source\\ \hline
 $K$& $\phantom{0}24.20 \pm 0.1$ & 0.5$^{\prime \prime}$ diameter, $K-F814$\\
 $K$& $\phantom{0}24.55 \pm 0.1$ & 0.5$^{\prime \prime}$ diameter, $K-model$\\
 $K$& $\phantom{0}23.40 \pm 0.13$ & 1.0$^{\prime \prime}$ diameter, $K-F814$\\
 $K$& $\phantom{0}23.85 \pm 0.13$ & 1.0$^{\prime \prime}$ diameter, $K-model$\\
   $K$& $\phantom{0}23.6\phantom{0} \pm 0.25$ & 1.0$^{\prime \prime}$ diameter, adopted average \\ 
 $K$& ${\phantom{0}23.4\phantom{0} \pm 0.25}$ & 1.0$^{\prime \prime}$ diameter, seeing corrected \\ 
$H-K$ & ${\phantom{02}1.4\phantom{0} \pm 0.35}$                                & 0.5$^{\prime \prime}$ diameter, seeing corrected\\
$I-K$ &${ >5.2}\phantom{02}$\phantom{,} $2\sigma$ & 0.5$^{\prime \prime}$ diameter, seeing corrected\\ \hline \hline
 $S_{\rm 450\mu m}$ & $<21$ \phantom{22.,} $3\sigma$\,mJy & SCUBA; Hughes et al. (1998)\\
 $S_{\rm 850\mu m}$ & $\phantom{<2}7.0 \pm 0.4$\,mJy & SCUBA; Hughes et al. (1998)\\
 $S_{\rm 1.3mm}$ & $\phantom{<2}2.2 \pm 0.3$\,mJy & IRAM; Downes et al. (1999)\\
 $S_{\rm 1.35mm}$ & $\phantom{<2}2.1 \pm 0.5$\,mJy & SCUBA; Hughes et al. (1998)\\ 
 $S_{\rm 8.4GHz}$ & $\phantom{<2}7.5 \pm 2.2$\,$\mu$Jy & E.Richards, priv. comm.\\
 $S_{\rm 1.4GHz}$ & $<23$ \phantom{22.,} 3$\sigma$\,$\mu$Jy & VLA; Richards (1999)\\ 
 $S_{\rm 1.4GHz}$ & $\phantom{12}16 \phantom{.5}\pm 4$ \phantom{2}\,$\mu$Jy & Merlin+VLA; this paper\\ 
\hline \hline
\end{tabular}
\end{table}

\begin{table}
\caption{\small Positions of the detections of 
HDF850.1 (see Fig. 4)}
\begin{tabular}{|l|l|l|} \hline 
Band & RA (J2000) & Dec (J2000)\\ \hline 
850\,${\rm \mu m}\,(1)$&12$^h$ 36$^m$ 52.32$^s\phantom{0} \pm 0.10^s$ & 
62$^\circ$ 12$^{\prime}$ 26.3$^{\prime \prime}\phantom{0} \pm 
0.7^{\prime \prime}$\\  
850\,${\rm \mu m}\,(2)$&12$^h$ 36$^m$ 52.22$^s\phantom{0} \pm 0.10^s$ & 
62$^\circ$ 12$^{\prime}$ 26.5$^{\prime \prime}\phantom{0} \pm 
0.7^{\prime \prime}$\\  
1.3\,mm     &12$^h$ 36$^m$ 51.98$^s\phantom{0} \pm 0.04^s$ & 
62$^\circ$ 12$^{\prime}$ 25.7$^{\prime \prime}\phantom{0} \pm 
0.3^{\prime \prime}$\\
1.4\,GHz    & 12$^h$ 36$^m$ 52.060$^s \pm 0.015^s$ & 
62$^\circ$ 12$^{\prime}$ 25.67$^{\prime \prime} \pm 
0.07^{\prime \prime}$\\
$K$       & 12$^h$ 36$^m$ 52.072$^s \pm 0.015^s$ & 
62$^\circ$ 12$^{\prime}$ 25.75$^{\prime \prime} \pm 
0.1^{\prime \prime}$\\
\hline 
\end{tabular}
\end{table}

\subsection{Properties of the elliptical galaxy 3-586.0}

As outlined in the introduction, the galaxy 3-586.0 is already known
to be a relatively quiescent, red elliptical with a fairly solid
estimated redshift of $z = 1.1 \pm 0.1$. As part of the analysis presented
here we have fitted a 2-dimensional de Vaucouleurs model to the new 
ground-based $K$-band, and existing HST $V$- and $I$-band images of 
this galaxy.
The results of this modelling are summarized in Table 3, where it can be seen
that the half-light radius $r_e$ depends on wavelength to a degree 
which is entirely consistent with the colour gradients exhibited by
other well-studied ellipticals of comparable size.

We have used these results to estimate the velocity dispersion of this 
elliptical galaxy from the $K$-band fundamental plane (Mobasher et al. 1999).
This was done in the following way. The observed $K$-band surface brightness
of 3-586.0, at the half-light radius $r_e$, is $\mu_e = 20.5$. 
To convert this to a present-day surface brightness we have made 
the standard correction for cosmological dimming, incorporating 
a $k$-correction based on a rest-frame colour of $J-K \simeq 1$, 
and assuming 
that between $z = 1$ and $z = 0$ this galaxy would be 
expected to dim
by $\Delta K \simeq 0.5$\,mag due to passive evolution. This latter assumption,
equivalent to assuming a minimum level of evolution (i.e. a high formation 
redshift), is justified by the fact that, with observed colours of 
$I-K \simeq 4$ and $H-K \simeq 1$, it is clear that 3-586.0 is 
already a very passive and 
well-evolved galaxy close to the red envelope of galaxy evolution at 
$z \simeq 1$. 

This calculation yields a rest-frame $K$-band surface brightness {\it at}
$r_e$ of $\mu_K = 18.7$, which converts to a mean surface brightness
internal to $r_e$ of $\langle \mu_K \rangle_{e} = 17.3$.

This number was then inserted into the relation between effective diameter
($A_e$), velocity dispersion ($\sigma_v$) and mean surface brightness 
($\langle \mu_K \rangle_{e}$) derived by Mobasher et al. for the Coma cluster:

\begin{equation}
 \log_{10}(A_e) = (1.38 \pm 0.26)\log_{10}(\sigma_v) + 
(0.3 \pm 0.02)\langle \mu_K 
\rangle_{e} - 7
\end{equation}

\noindent
To obtain $\sigma_v$ from this relation we calculated $A_e$ for 3-586.0 by 
first converting the geometric average value of $r_e$ given in Table 3 to
the major axis value, doubling this and then transforming the result
to the redshift of the Coma cluster ($z = 0.0231$). This yields a 
value of $A_e = 14.8$\,arcsec.

The resulting estimate for the velocity dispersion of 3-586.0 is
$\sigma_v = 146 \kms$, with an rms uncertainty of 
$\pm 29 \kms$ which we have calculated directly from the
data for the Coma cluster provided by Mobasher et al.

This new estimate for the velocity dispersion of 3-586.0 is somewhat lower
than the value previously 
estimated by Hogg et al. (1996). However, this earlier estimate
was based purely on the Faber-Jackson relation, and did not allow for the
effects of passive evolution.

\begin{table*}
\caption{\small Properties of the elliptical galaxy 3-586.0. Magnitudes
were determined through an aperture of diameter 3\,arcsec. Half-light
radii assume $H_0 = 70~\kms\,{\rm Mpc^{-1}}$, $\Omega_m = 0.3$,
$\Omega_{\Lambda} = 0.7$. Velocity dispersion was estimated from 
the $K$-band fundamental plane (see section 5.2).} 
\begin{tabular}{|c|c|c|l|l|c|} \hline \hline
RA (J2000) & Dec (J2000) & $z$ & Magnitude & Scalength $r_e$ & velocity dispersion\\ \hline 
12$^h$ 36$^m$ 52.101$^s \pm 0.014^s$ & 
62$^\circ$ 12$^{\prime}$ 26.27$^{\prime \prime} \pm 
0.1^{\prime \prime}$ & $1.1 \pm 0.1$ & $K = 19.39 \pm 0.03$ & 3.0 kpc (at $K$) &
$146 \pm 29$ $\kms$\\
&&&$H = 20.40 \pm 0.05$&3.0 kpc (at $H$) &\\
&&&$I = 23.40 \pm 0.05$&3.3 kpc (at $I$) &\\
&&&                    &4.0 kpc (at $V$) &\\
\hline \hline
\end{tabular}
\end{table*}

\subsection{The estimated redshift of the sub--mm source}

\begin{figure}

\vspace*{13.5cm}
\includegraphics{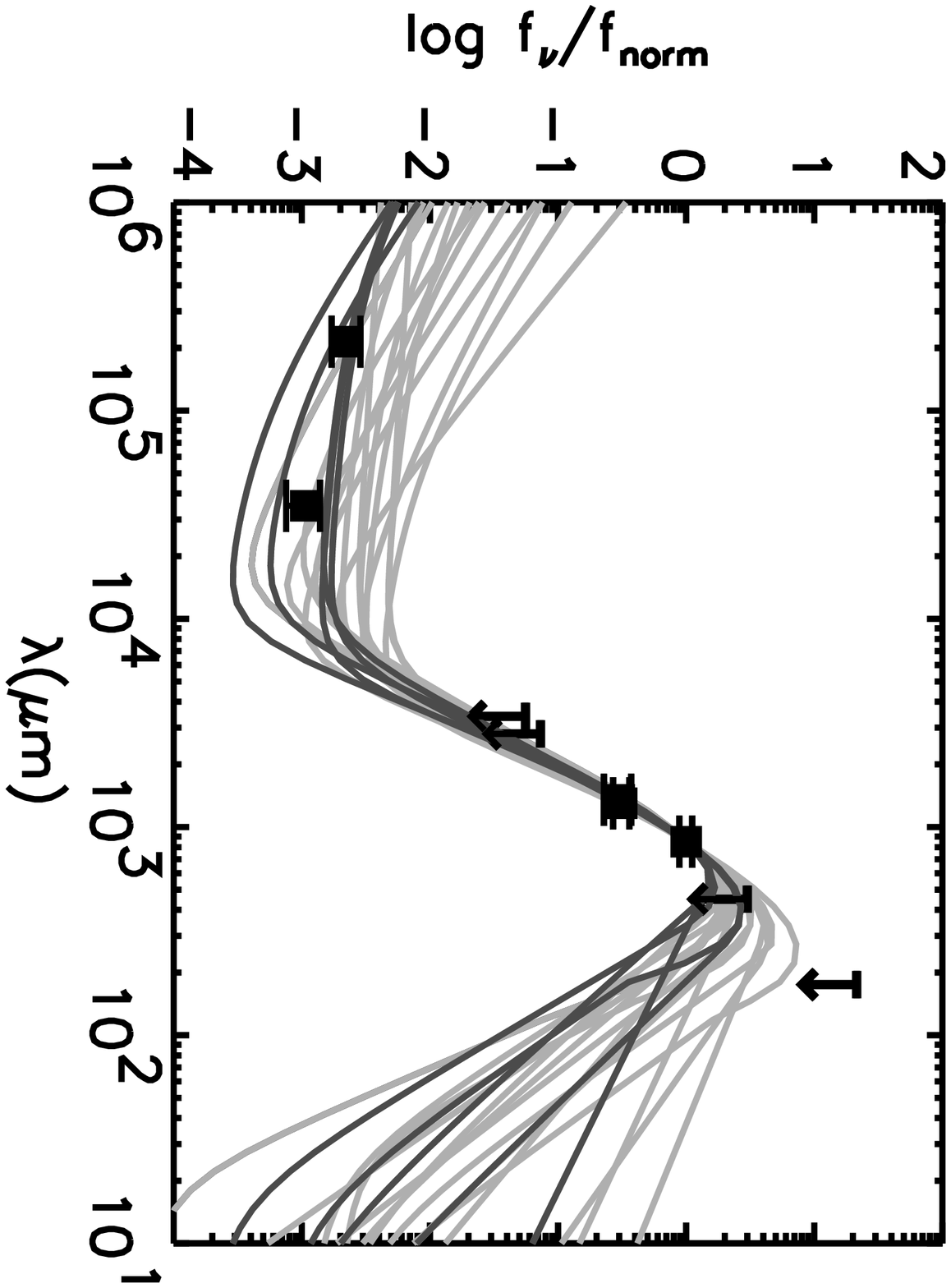}
\includegraphics{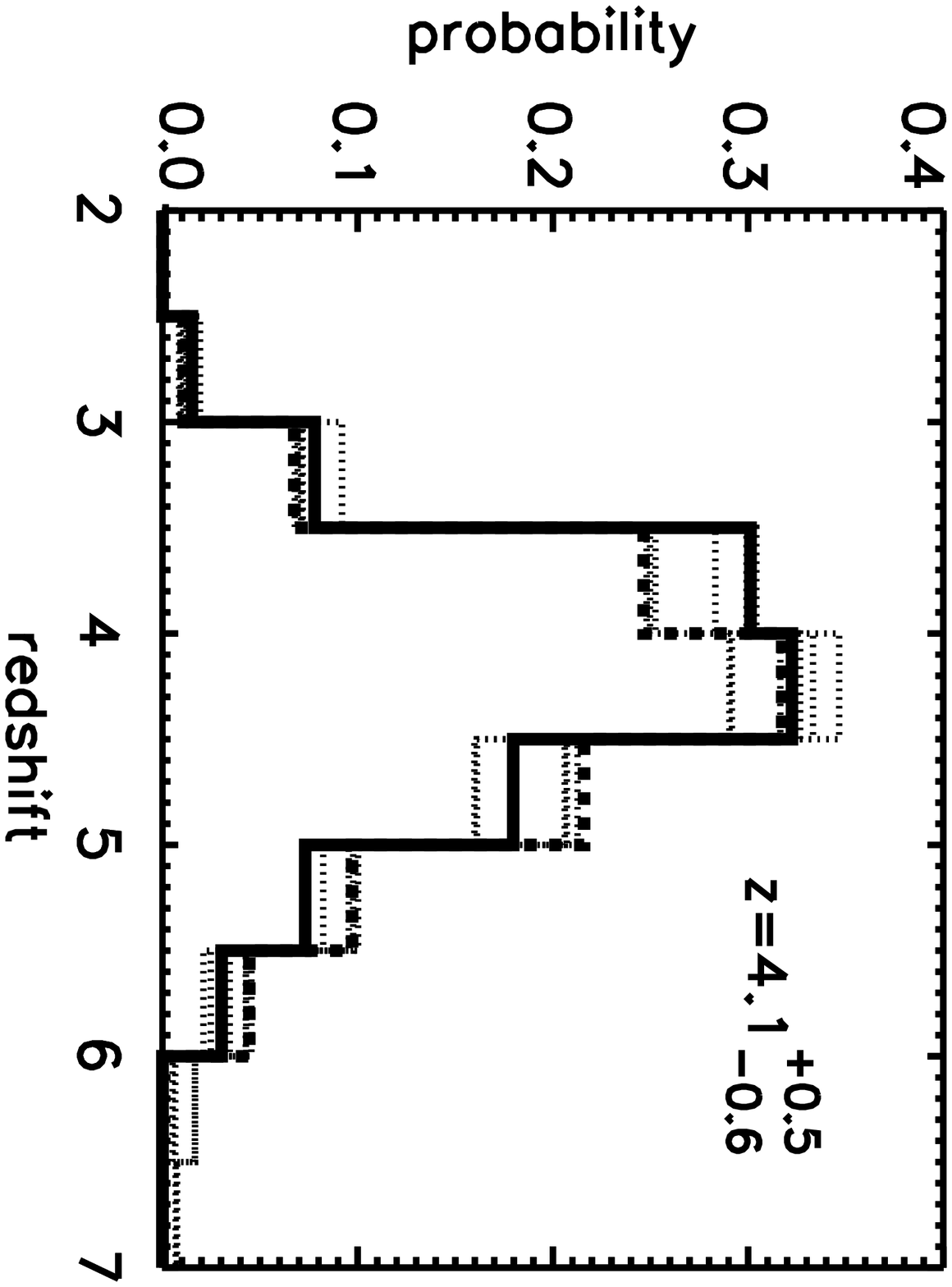}
\caption{\small{Results from the Monte Carlo photometric redshift
estimation.
Upper panel: SED of HDF850.1 normalized to 850\,$\mu$m, 
where squares are detections 
with superimposed 1-$\sigma$ error bars, and arrows indicate 
3-$\sigma$ upper limits. The lines represent the SEDs of local
starbursts, ULIRGs and AGN used in the calculation,
redshifted to $z=4.1$. Dark grey lines 
are SEDs compatible at a 3-$\sigma$ level with the SED of HDF850.1, 
and light grey lines are SEDs which are not compatible within the
3-$\sigma$ level.
Lower panel: Redshift probability distributions derived for
HDF850.1. Different lines correspond to different adopted
evolutionary models and amplification factors (see Aretxaga et al.
2002 for details). The thick solid line represents an unlensed
scenario for a given evolutionary model, and the thick dashed line
represents the corresponding lensed scenario.
The mode of the redshift distribution and the 
68\% confidence
interval are indicated within the panel for the unlensed model. 
The lensed scenario yields an indistinguishable result.}}
\end{figure}

As discussed in Sections 2 and 3,
the SED of HDF850.1 is well sampled from radio to FIR
wavelengths. The upper panel of Fig. 5
shows the observed long-wavelength SED of HDF850.1, normalized to
its flux density at 850\,$\mu$m.
The shape of this SED can be used to determine 
the redshift probability distribution
of the source. A Monte Carlo photometric redshift 
method has been developed (Hughes et al. 2002; Aretxaga et al. 2002)
to also take into account constraining 
prior information such as the number counts of sub-mm galaxies,
the favoured evolutionary model of the sub-mm population,
and the amplification and clustering properties of a certain field.
We assume that the SEDs of sub-mm galaxies are well represented
by 20 SEDs of local starbursts, ULIRGs and AGN, all of which are
well sampled in the radio--FIR regime to allow the reliable 
fitting of physically-motivated functional relationships. 
 The Monte Carlo yields
the probability of producing the colours and flux densities 
of the sub-mm galaxy under study
at any given redshift, 
and thus provides us with the whole redshift probability distribution 
and not  
with just the first and second moment of the distribution, as popular
maximum-likelihood techniques do.

This Monte Carlo photometric redshift technique 
places the most
likely redshift of HDF850.1 at $z=4.1$
with a 68\% confidence interval of $3.5\le z \le 4.6$ and a 90\%
confidence interval of $3.0 \le z \le 5.1$.
The redshift probability
distribution is clearly consistent with earlier claims
that locate HDF850.1 at 
$z\ge 2.5$ (Hughes et al. 1998; Downes et al. 1999).
The lower panel of Fig. 5 shows the redshift probability 
distributions found
under different assumptions (details about the calculations 
are given in Aretxaga et al. 2002).
 Different lines correspond to different adopted
evolutionary models of the sub-mm population and possible 
lensing amplifications.
The thick solid line corresponds to an unlensed
 scenario and the thick dashed line to a lensed scenario
where HDF850.1 is amplified by a factor of 3 (see section 6.2). 
The results are almost independent of the adopted 
evolutionary model of the 
sub-mm population and the lensing amplification of HDF850.1
considered. 
Lensed models
produce a minor transfer of probability from redshifts $z<4$ to 
redshifts $z>4.5$, as the likelihood of detecting fainter, 
higher-redshift sources with compatible colours is increased. 
However, the most probable redshift in the lensed scenarios is still 
basically the same, i.e. $z=4.2$.

\section{Discussion}

\subsection{Robustness of the sub-mm source identification HDF850.1K}

\subsubsection{Statistical association with the IRAM source}

To estimate the statistical confidence with which this 
newly-discovered faint ERO can by identified with the SCUBA source,
we first ignore the new radio information and simply add 
HDF850.1K to the list of possible identifications 
for the 1.3\,mm detection of HDF850.1 considered by Downes et al. (1999).
As illustrated in Fig. 4b, the distance from the IRAM position 
to HDF850.1K
is 0.65\,arcsec. The number density of galaxies with $K < 
23.5$ is $\simeq 0.02$\,arcsec$^{-2}$ (Maihara et al. 2001).
Consequently, the raw Poisson probability of finding HDF850.1K 
so close to the IRAM position is $P_0 = 0.03$. However, as
discussed by Downes et al. (1986) (see also Dunlop et al. 1989,
and Serjeant et al. 2002),
this number needs to be corrected for the different ways that objects 
with this a posteriori 
Poisson probability could be found down to the limiting
magnitude of the available data. Using the corrected number counts from
Table 3 of Maihara et al., the cumulative surface density of objects
down to $K = 25.5$ is 0.07\,arcsec$^{-2}$. So, for an 
adopted search radius around the IRAM source of 1\,arcsec, the corrected 
probability that this is a chance coincidence is $P = 0.09$.

It is interesting to consider how this compares with the statistical
probability that the elliptical 3-586.0 lies so close to the IRAM 
source by chance. In this case, using the $I$-band HST data, we get 
a range of values depending on whether we use the catalogue of 
Williams et al. 1996 (W96), the SExtractor catalogue (N98), or the shallower 
Barger et al. (1999a) catalogue (B99). The raw Poisson probabilities 
of this coincidence calculated from these three 
different catalogues are $P_0 = 0.07$ (W96), 0.03 (N98) or 0.01 (B99).
If we again adopt a search radius of 1\,arcsec around the IRAM source,
the corrected probabilities become $P = 0.16$, 0.09 or 0.03 respectively.
The probabilities for 3-593.1 are somewhat higher and, as already noted, 
3-577.0 lies outwith any reasonable search radius from the IRAM position.

Thus, ignoring the radio data we can conclude that the probability that
HDF850.1K lies so close to the IRAM source position by chance is
$\simeq 5$\%, and that this 2-$\sigma$ result is comparable to the statistical
significance of the IRAM+3-586.0 association. If, as has been argued
above, 3-586.0 can be rejected as a potential SCUBA identification on
the grounds of estimated redshift, we can conclude that HDF850.1K
is the most likely counterpart for the IRAM source as revealed
by all optical and near-infrared imaging undertaken to date.

Qualitatively, this identification seems all the more likely because
it is an ERO ($I-K > 5.2$, $H-K \simeq 1.4$), given that several of the most 
secure SCUBA identifications have comparable colours (e.g. Lutz et al.
2001; Smail et al. 1999; Dey et al. 1999; Ivison et al. 2000, 2001). 
However, factoring this observation into
the statistical calculation is difficult because the surface density
of EROs is not properly determined at this depth. Specifically,
while $K$-band number counts reaching $K=23$ have been published in three
papers (Maihara et al., Bershady et al. 1998 and Moustakas et al. 1997)
in only one of these are deep $I$-band data also presented 
(Moustakas et al. 1997), and in this case the authors 
find no galaxy with $I-K > 5$ and $K > 23$ in a 2\,arcmin$^2$ field.

Alexander et al. (2001) define `very red objects' (VRO) 
as having $I-K > 4$
and report a cumulative number density of 1500\,($+250, -250$)\,degree$^{-2}$ 
to $HK^{\prime} = 20.4$, and
6100\,($+2800, -2000$)\,degree$^{-2}$
to $K = 22$. So, down to $K=22$ we can calculate that 
the probability of finding a `VRO' 0.65 arcsec from the IRAM position by
chance (even adopting a source density of 8900\,degree$^{-2}$) is
$P_0 = 0.001$. Clearly this number will be increased by going to increased
depth in $K$, but it will also decrease if the redder threshold
$I-K > 5$ appropriate to the current study is adopted.
In fact, these two factors will roughly cancel out, 
as can be seen from an extrapolation
of the results of Smith et al. (2002), 
who estimate the number densities
of EROs (in $R-K \geq 5.3$ and $R-K \geq 6.0$ subsamples) 
down to K=21.5.
Extrapolating the results of their redder subsample, 
as appropriate for 
HDF850.1K, yields a very rough estimate of $\sim$0.001 
ERO arcsec$^{-2}$
at K=23.5, which translates to a probability of 
$P_0 \sim 0.001$ that
such an object would be found 0.65 arcsec from the 
IRAM position by
chance.

\subsubsection{Statistical association with the new 1.4-GHz detection}

The statistical probabilities calculated above are already 
reasonably compelling. However, we have not yet made use  
of the new 1.4-GHz MERLIN+VLA detection to refine the expected position
and search radius for the optical/IR counterpart.

To do this we first calculate the probability that this 
1.4-GHz MERLIN+VLA source is indeed the same source as detected by 
IRAM at 1.3\,mm. In fact, since the angular 
separation of these two sources is only 0.56\,arcsec
(see Fig. 4b), and the surface density
of radio sources down to $S_{\rm 1.4GHz} > 17$\,${\rm \mu Jy}$ is only 
$\simeq 0.0009$\,arcsec$^{-2}$, the Poisson probability of such a chance
coincidence transpires to be only $P_0 = 0.0009$.

We can therefore safely assume that the 1.4-GHz source is indeed the 
mm/sub-mm source, and therefore 
that the MERLIN+VLA 1.4-GHz detection 
offers the most accurate available position
for HDF850.1. If we now recalculate the probability 
that HDF850.1K is a chance coincidence with HDF850.1
we find a raw Poisson probability of $P_0 = 0.0009$, corrected
to $P = 0.008$ if we adopt a 3-$\sigma$ search radius of 0.3\,arcsec.
This is a compellingly small number, which would only become smaller if 
we were able to factor in reliably 
the prior probability that the object is an ERO.

We conclude by noting that the 3-$\sigma$ search radius from the MERLIN+VLA
position does not now extend to include the centroid of 3-586.0
(see Fig. 4), and 
so this object can now be excluded as a possible identification without
recourse to arguments based on redshift estimates for both sources.
Of course, the fact that 3-586.0 has eventually 
turned out {\it not} to be a feasible identification on astrometric grounds
in a sense lends credence to the pre-existing redshift-estimate argument.

However, despite its rejection as a possible identification, it still
remains the case that 3-586.0 lies surprisingly close to HDF850.1 on 
the sky (at the 2-$\sigma$ level), a statistical result which merits 
a physical explanation. The most likely explanation is that HDF850.1
has been gravitationally lensed by 3-586.0,
a possibility which we explore and quantify below in section
6.2.

\subsubsection{Optical-IR constraints on the redshift of the 
HDF850.1K}

\begin{figure}
\vspace*{14.5cm}
\includegraphics{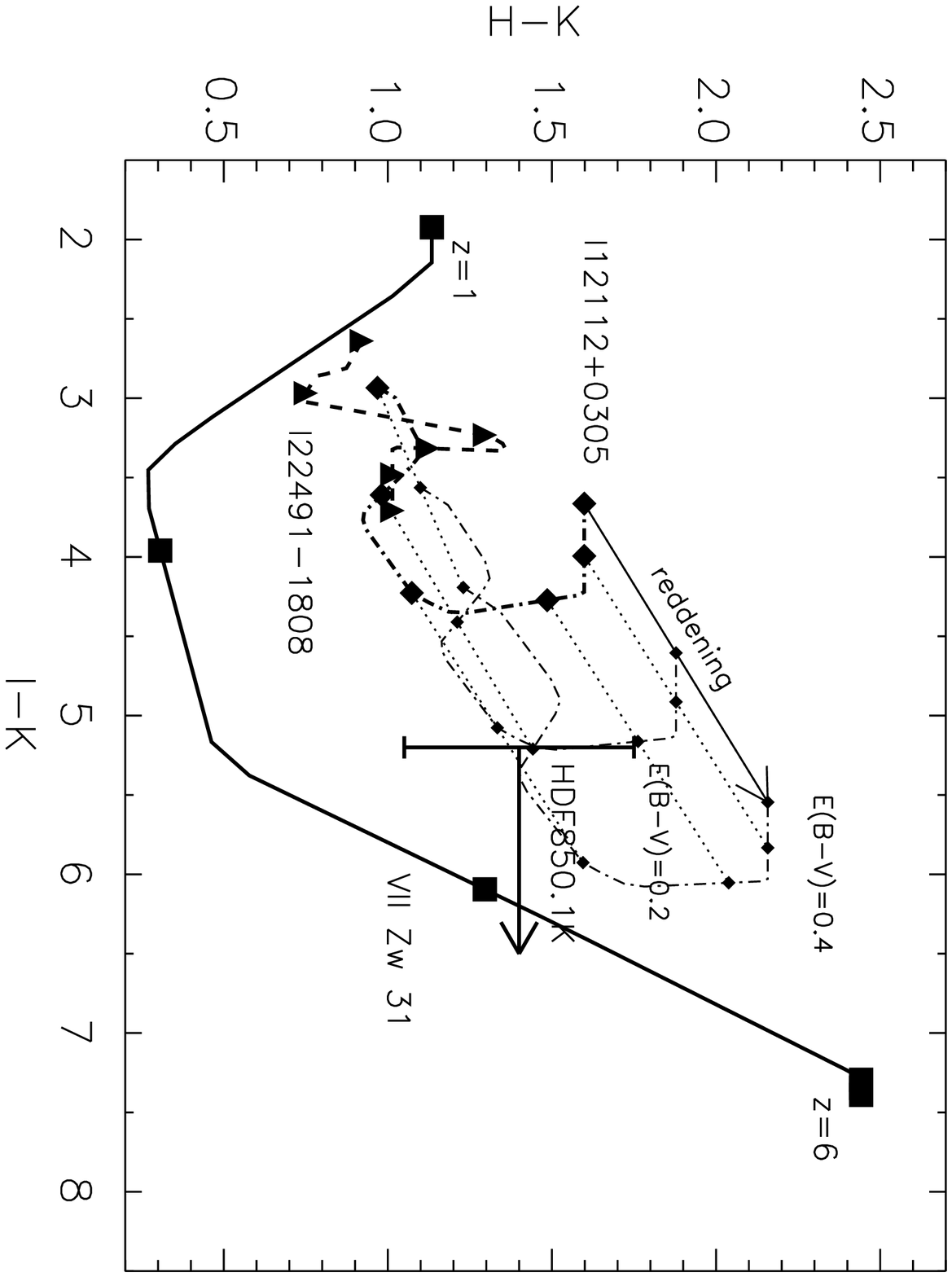}
\includegraphics{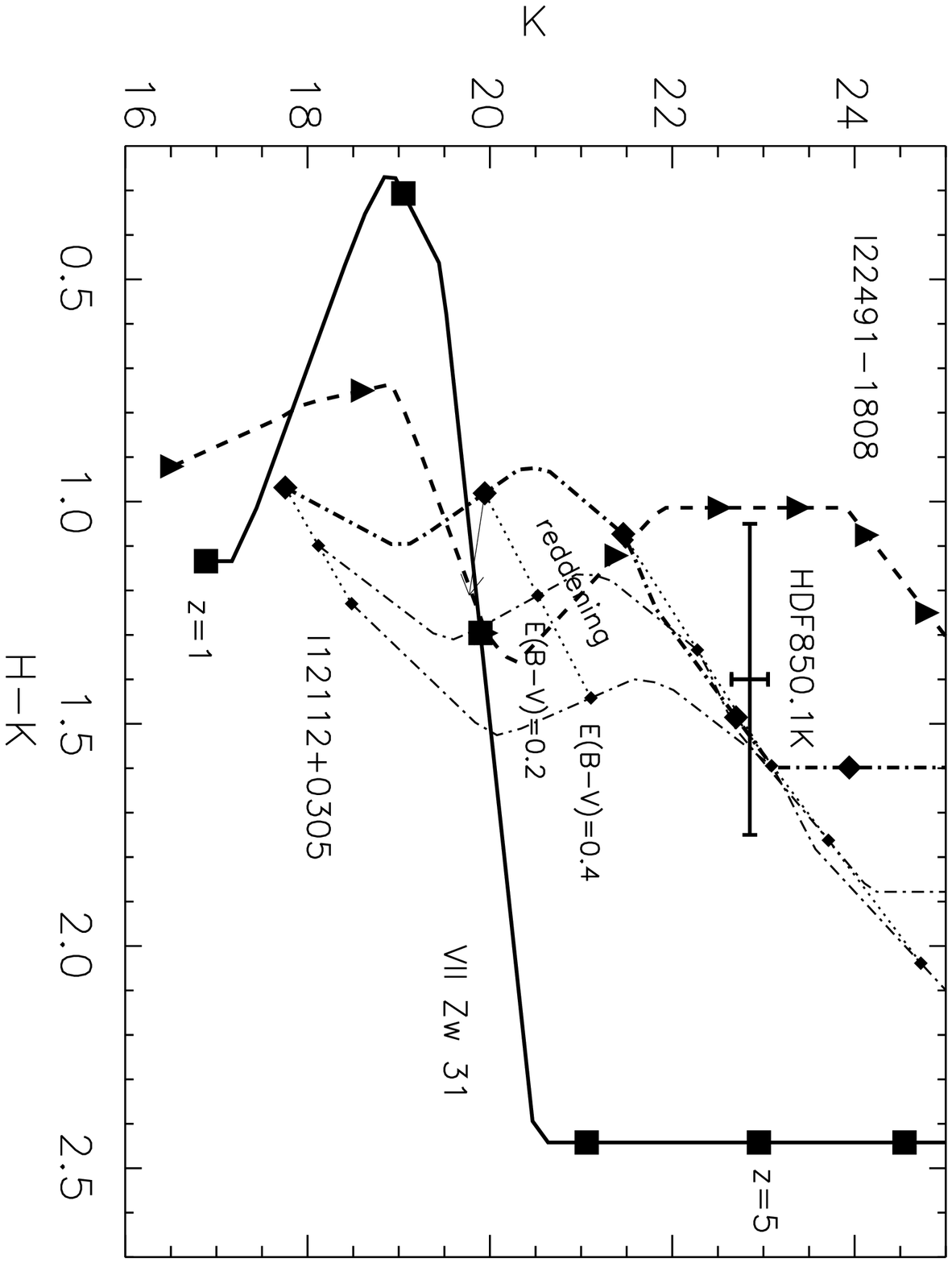}
\caption{\small{$H-K$ vs. $I-K$, and $K$ vs. $H-K$ diagrams. 
showing the predicted tracks  of 5-mJy 1.2\,mm sources
with increasing redshift, taking 
the local ULIRGs I12112+0305, I22491$-$1808 and VII Zw 31
(thick lines) as analogues. 
The symbols represent the position in the track 
of each galaxy
at $\Delta z=1$ steps, starting from $z=1$ at the bottom left of each figure. 
The effect
of an increase in reddening is shown by the parallel curves to the right of 
I22491$-$1808, in thin lines and small symbols. HDF850.1K appears 
near the centre of each figure, with its $H-K$ and $I-K$
colours, and (appropriately scaled)
$K$ magnitude indicated by 1$\sigma$ error bars.}
}
\end{figure}

Having established that HDF850.1K is statistically highly likely to be the
true host galaxy of the sub-mm source, it remains to consider
whether, unlike 3-586.0, its optical-infrared photometric 
redshift estimate
is at least consistent with that derived 
for the sub-mm source from the longer-wavelength data in section 5.3.

Given that the available information on HDF850.1K 
consists of only a $K$ magnitude, an approximate $H-K$ 
colour, and a limit on $I-K$ colour, it is obviously
impossible to make an accurate estimate of its redshift. However,
it is still worth exploring what constraints can be placed on its redshift 
under the assumption that its intrinsic properties are not unlike
those of low-redshift ULIRGs. This assumption is of course debatable,
and indeed there is some evidence that the near-infrared 
to ultraviolet properties of local ULIRGs do not 
quite correspond
to the red colours of the sub-mm population (local ULIRGs are 
in general bluer than the EROs associated with 
sub-mm galaxies; Dannerbauer et al. 2002).
Nevertheless,
in the absence of a good set of UV--optical--IR templates to 
determine the photometric redshift of HDF850.1K, we have used the
local sample of ULIRGs to look for constraints on its redshift,
independently of the redshift associated with the radio--mm--FIR
source. 

The upper panel of Fig.~6 shows the $H-K$ vs. $I-K$ colours of 
three local ULIRGs with available UV--optical colours 
(Trentham et al. 1999) when placed at different redshifts.
At  $z>4.7$, the predicted $I$ magnitude is calculated based on 
extrapolations from the rest-frame
1450\AA\ measurements, the shortest-wavelength at which these ULIRGs have been
detected, and thus this redshift regime should be interpreted with caution.
The colour offset
produced by increasing amounts of reddening E($B-V$)=0.2, 0.4
according to a typical starburst extinction law (Calzetti et al. 2000)
are shown for I12112$+$0305, and are similar for the other ULIRGs.
The colours of HDH850.1K can be reproduced within the 1$\sigma$ confidence
interval by
VII~Zw~31 at $z=2.8-3.2$ ($2.6-6.0$ within 3$\sigma$), 
by I12112$+$0305 at $z=3.3-4.0$ allowing for a E($B-V$)=0.2 increased 
extinction ($z=3.3-4.5$ within 3$\sigma$), and by 
I22491$-$1808 at $z=3.5-4.0$ allowing for a E($B-V$)=0.4 increased extinction
($z=4.7-6.0$ within 3$\sigma$).
Increasing amounts of extinction over these values can also accommodate 
lower redshift ranges, as shown by the parallel tracks of 
I12112$+$0305 in Fig.~6.
Similarly, de-reddening
values of E$(B-V) \le$ 0.5
can accommodate higher redshift intervals ($z<6$) for
VII~Zw~31.

Recently, Dannerbauer et al. (2002) have produced a diagnostic
diagram to estimate the redshifts of mm-galaxies
based on the $K$ magnitude of their IR counterparts, 
taking a local sample of ULIRGs as templates. 
Their method provides a complementary constraint on the redshift 
interval compatible with the optical--IR observations of HDF850.1K.
The lower-panel of Fig.~6 shows the  $K$ vs $H-K$ magnitude--colour
relationship that the local ULIRGs sample describes 
when located at different redshifts.
Following Dannerbauer et al. (2002), the $K$-band magnitudes are derived 
from scaled UV--optical--IR--mm SEDs that reproduce 5mJy sources at 1.2mm,
and HDF850.1K is, therefore, accordingly scaled to $K=22.85\pm0.20$.

Disregarding reddening corrections, the only ULIRG compatible
with the $H-K$, $I-K$ colours and $K$-band magnitude   
of HDF850.1K within a 3$\sigma$ confidence level  
is VII~Zw~31 at $z=4.7-5.2$. However, a small increase E($B-V$)= 0.2 
in the amount of reddening of I12112$+$0305 can reproduce the colours
of HDF850.1K within a 1$\sigma$ confidence interval at $z=3.3-3.5$,
and  within a 3$\sigma$ confidence interval at $z=3.0-3.8$.
For I22491$-$1808, the same is true using additional extinction corrections
E($B-V$)= 0.4, at $z\approx 3.8$ within a 1$\sigma$ interval, and
at $z\approx 3.4-4.0$ within a 3$\sigma$ interval.

In summary, if placed at 
$z=3.0-5.2$, the template local ULIRGs 
could reproduce the $H-K$, $I-K$ colours and $K$ magnitudes  
of HDF850.1K
within the 3$\sigma$ confidence interval. The probability that 
the redshift of the sub-mm source
is in the $z=3.0-5.2$ regime is 90\%  based on its 
radio--sub-mm--FIR colours (section~5.3.) . 
Therefore, although the derivation of a reliable optical--IR 
photometric redshift
for HDF850.1K is not possible given the present optical-infrared data,
this analysis does at least show that these are completely 
consistent with the redshift
interval derived in section~5.3. 

\subsection{Gravitational lensing}

\begin{figure}
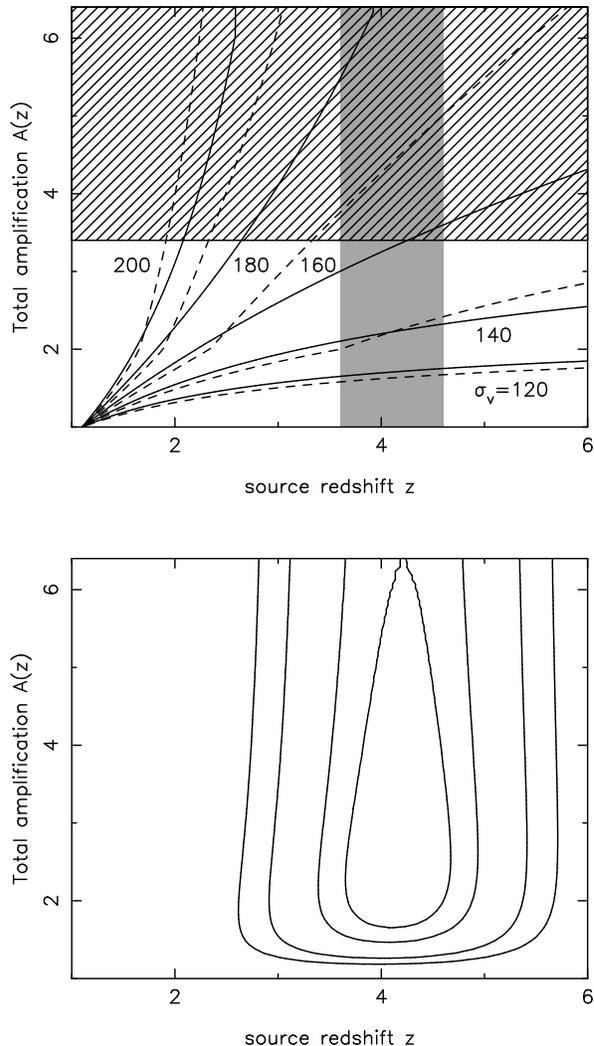

\vspace*{14.5cm}
\includegraphics{fig7a.ps}
\includegraphics{fig7b.ps}

\caption{\small The effect of gravitational lensing on HDF850.1.
The upper panel shows predicted total lensing 
magnification as a function of source redshift for different assumed 
values of the velocity dispersion ($\sigma_v$) of the lensing 
$z \simeq 1$ elliptical 3-586.0. 
The dashed lines show the case of a singular isothermal sphere;
the solid lines show the more realistic case in which the mass
model has a core (see text).
The 2-$\sigma$ limit on any counter-image is a
flux density 0.26 times that of the main image; this
corresponds to a total amplification $A<3.4$ for the singular
model (hatched area), or $A<6.4$ for the model with a core. 
If the photometric redshift indication
for HDF850.1 is at all realistic, 
the velocity dispersion cannot exceed about $160\kms$,
consistent with the 
velocity dispersion of $\sigma_v = 146 \pm 29 \kms$
deduced independently from the infrared fundamental plane (see Section 5.2
and Table 3). The lower panel 
shows likelihood contours for the total lensing magnification
of HDF850.1 and its redshift, taking account of the
estimated velocity dispersion for the foreground
elliptical, the limits on any counter-image, plus the photometric 
redshift and its uncertainty. Likelihood contours are plotted 
at the usual positions
for one-parameter 68\% confidence and two-parameter
confidence values of 68\%, 95\%, and 99\% (i.e.
offsets in $\ln{\cal L}$ of 0.5, 1.15, 3.0 and 4.6).
We see that it seems inevitable that HDF850.1
has been subject to significant lensing amplification:
a 95\% confidence lower limit of a factor 1.7.
Note that we cannot rule out the alternative possibility of stronger
on-axis lensing, in which a slight ellipticity of the lens would
break the symmetry between main and counter-image.
This would require a more massive lens, with $\sigma_v \simeq 200 \kms$.
}
\end{figure}

\subsubsection{The HDF850.1 + 3-586.0 system}

One striking aspect of the new proposed identification of
HDF850.1 is its proximity to the elliptical 3-586.0.
With a low photometric redshift ($z = 1.1\pm0.1)$, this was considered
a poor candidate identification by Hughes et al. (1998). 
Hughes et al. nevertheless recognized that a $z \simeq 1$ elliptical
such as 3-586.0 had a low probability of aligning
with a SCUBA source by chance, and they proposed that gravitational
lensing by the foreground elliptical could have enhanced
this probability. We now revisit this possibility in the
light of our improved knowledge of the identification of
HDF850.1 and the properties of 3-586.0.

The main features of the putative lens-source system are as follows.
The observed separation of lens and main image is 0.55\,arcsec, but there is
no suggestion of a counter-image. This statement does not assume that
any counter-image would be exactly opposite the main image; 
Fig. 2 shows that there is no other significant image in
the general vicinity. Based on the noise within 0.5-arcsec
diameter photometric apertures on the $K^{\prime}$ image, 
we adopt a 2-$\sigma$ limit for a counter-image
of $K > 26$, which implies an upper limit of 0.26 for the flux ratio
between any counter-image and the main image. This lack of
evidence for strong lensing constrains a combination of the mass of 3-586.0
and the redshift of HDF850.1; in the following discussion, we assume
a standard $\Omega=0.3$, $k=0$ geometry.

The simplest lens model to consider for 3-586.0 is a
singular isothermal sphere, for which the only
parameter is the angular radius of the Einstein ring:
\begin{equation}
\theta_{\japsub E} =
 \left({\sigma_v \over 186 \kms}\right)^2 \; {D_{\japsub LS} \over D_{\japsub S}}\quad{\rm arcsec}.
\end{equation}
For an observed separation $\theta$ between lens and main image,
multiple imaging is predicted if $\theta < 2\theta_{\japsub E}$, with total
amplification $A=2\theta_{\japsub E}/(\theta-\theta_{\japsub E})$, and a
flux ratio between counter-image and main image of $2\theta_{\japsub E}/\theta -1$.
The lack of an obvious counter-image suggests that we cannot be very
far into the strong-lensing regime; the 2-$\sigma$ limit is a counter-image
with flux density 0.26 times that of the main image, which
corresponds to a total amplification $A=3.4$. Together with the
observed radius of the main image at 0.55~arcsec from the centre of 3-586.0,
this yields the following constraint on the Einstein radius:
\begin{equation}
\theta_{\japsub E} < 0.35\; {\rm arcsec}; \quad A_{\rm max}=3.4.
\end{equation}
This allows a redshift-dependent
limit to be set on the velocity dispersion of 3-586.0, as shown
in Fig. 7a. If the photometric redshift indication
for HDF850.1 is at all realistic, 
the velocity dispersion of 3-586.0 cannot exceed about $160\kms$
in this model.

A more realistic model will have a finite core to the mass distribution.
We assume that 3-586.0 is baryon-dominated in the centre,
so that the observed $r^{1/4}$ profile gives the bend-angle
profile directly, up to a free $M/L$. With this assumption
the bend angle has a broad peak at 0.9~arcsec radius, falling to
half this value at 0.15~arcsec. We connect this to the isothermal
sphere model by assuming that the bend angle remains constant
beyond 0.9~arcsec, and we use this asymptotic value to give an
effective velocity dispersion to the lens, which is most
conveniently quoted as the Einstein radius that
corresponds to a singular
model with the same asymptotic bend angle.
For this non-singular profile, the onset of multiple lensing is more
abrupt: additional images first form
with divergent amplification  via caustic-crossing,
at which point the amplification of the main image is
$A=6.4$. The limit on the lens properties now changes to
\begin{equation}
\theta_{\japsub E} < 0.47\; {\rm arcsec}; \quad A_{\rm max}=6.4.
\end{equation}
This permits a slightly more massive lens: up to $\sigma_v = 178\kms$
for $z_{\japsub S} = 4.1$, as compared to a limit of
$\sigma_v = 154\kms$ for the singular model with this source redshift.
Given the velocity dispersion estimate of $\sigma_v = 146 \pm 29 \kms$
for 3-586.0, inferred from the $K$-band fundamental plane
in Table~3, it seems that 3-586.0 must be causing
significant lensing amplification.
This is shown in Fig. 7b, which evaluates the likelihood over
the plane of source redshift and total amplification
(assuming the non-singular model), using
the estimate of the
velocity dispersion (assuming the error distribution to be
Gaussian in $\ln\sigma$) and the photometric source redshift 
estimate of $z = 4.1\pm0.5$  (assuming the error distribution
to be Gaussian in $z$). Integrating the likelihood values
up to $A_{\rm max}=6.4$ yields a median amplification of
3.6, with a 95\% confidence lower limit of 1.7.

For a more massive lens, constrained to yield one image at 0.55~arcsec
radius, significant counter-images are generally predicted. 
There are however two ways of evading the conclusion that the maximum
lensing amplification is 6.4. The first is to consider the possibility that
the observed image is in fact not the principal image: if the true
position of HDF850.1 is actually on the opposite side of 
3-586.0, the observed image could be the result of a merging
pair of secondary images produced when the source crosses the
outer caustic. The amplification for this situation can be
divergent. This seems improbable, however, since the lens would need
to be unrealistically massive in order to produce secondary
images at 0.55~arcsec ($\theta_{\japsub E} = 2.8$~arcsec).
A more realistic alternative with extreme amplification results
if we relax the assumption of a circularly symmetric lens, in accord with
the measured axial ratio of 1.28 for 3-586.0. This has little
effect on the lensing properties at moderate amplifications, but
greatly changes the high-amplification limit. An inner caustic
opens up, and it is possible to achieve divergent amplification
with large image flux ratios if the source is placed close to
this caustic. For this option, the lens mass has to be such that
the observed critical line passes through the image radius
of 0.55~arcsec. For a circular lens, this would require
$\theta_{\japsub E}=0.57$~arcsec, or $\sigma_v = 198\kms$
for $z_{\japsub S} = 4.1$.

Finally, we should consider whether it is legitimate to model
the lens as a single component. If 3-586.0 were to lie in a rich cluster with a
constant-density core just below the critical
surface density, strong lensing effects could arise with a much
less massive galaxy -- e.g. the case of cB58 (e.g. Seitz et al. 1998),
which has some similarities to the present situation.
However, cB58 was found towards a
pre-selected X-ray luminous cluster; even at $z=1.1$,
it is clear that nothing remotely as massive exists around 3-586.0. 
Assuming a source redshift of 4.1, the critical surface density
at 3-586.0 is $7.2\times 10^{14}\, h \, M_\odot\, {\rm Mpc}^{-2}$;
the most massive known clusters at $z=0.8$ have
surface densities similar in magnitude to the the critical
value of interest here (Clowe et al. 1998), but these are rare
systems, which would be extremely prominent on data of the HDF depth.

Such systems would also be very easily
detected at X-ray wavelengths in the deep Chandra imaging of this field.
In fact, the 1\,Msec Chandra image centred on the HDF
revealed 6 extended sources in the full $16 \times 16$\,arcmin
field (Bauer et al. 2002). These can be used to obtain a 
conservative upper limit on any cluster emission in the vicinity 
of HDF850.1, which lies in the most sensitive part of the X-ray 
image. The two faintest extended sources from 
Bauer et al. (2002) have an X-ray
flux density of $\simeq 3 \times 10^{-19}\,{\rm W m^{-2}}$ 
($0.5-8$\,keV). There is no evidence for any
X-ray emission in the vicinity of HDF850.1 which, 
assuming $z \simeq 1$ for a high-redshift cluster containing 
3-586.0, leads to an upper limit on 
cluster X-ray luminosity of $L_X < 2 \times 10^{35}\,{\rm W}$.
This corresponds to the X-ray luminosity of a weak group, 
$\simeq 100$ times fainter
than the ROSAT and EMSS $z \simeq 1$ clusters discussed
by Clowe et al. (1998).

In summary, there are two possibilities regarding 
gravitational lensing of HDF850.1, assuming it to lie
at $z=4.1$: (i) A lens
with $\sigma_v < 178\kms$ would imply that there is no
multiple imaging, and that the total amplification is
moderate ($A\simeq 3$); (ii) for $\sigma_v$ close to $200\kms$, 
there is also the possibility of extreme nearly on-axis lensing,
in which the amplification is not constrained, and could plausibly be 
more than 10 times larger.
The velocity dispersion estimate of $\sigma_v = 146 \pm 29 \kms$
for 3-586.0 favours the former model, but does not completely rule
out the on-axis case.
One may suspect that $K\simeq 23.5$ is already so faint
that cB58-like amplifications of order 30 are implausible, but in the
absence of other evidence it is best to consider the
a priori probabilities of these outcomes, as discussed below.

\subsubsection{Estimated prevalence of lensing within 
the sub-mm source population}

What are the implications of HDF850.1 for the occurrence of lensing
in other SCUBA sources? It is important to understand whether this
is an isolated rare event, or if lensing needs to be taken into
account in interpreting the data on all bright SCUBA sources
(see also Chapman et al.\ 2002).

The expected amplification distribution is $P(>A) = \tau/A^2$,
so we need an estimate of the lensing optical depth, $\tau(z)$.
For a source at $z=4.1$, this is $\tau=0.12$ on the assumption that
the galaxy population at $z\simeq 1$ is unchanged compared to $z=0$
(e.g. Peacock 1982). Note that the optical depth is not sensitive
to the exact lens model, and depends mainly on the total density
of mass in systems of above critical surface density.
The neglect of evolution means that $\tau=0.12$ is probably an
overestimate. Figures about a factor of
4 lower are obtained from calculations based on evolving
dark-matter haloes (e.g. Perrotta et al. 2002). However, because
these estimates neglect halo substructure, they are almost
certainly too low. The correct figure is unlikely to be far from
$\tau=0.1$, so the intrinsic probability of strong ($A>2$)
lensing is only about 2\%.  This is boosted by amplification bias:
if the slope of the counts is $dN/d\ln S \propto S^{-\beta}$, the
biased lensing probability is

\begin{equation}
P(>A) = {2\over 2-\beta}\; {\tau \over A^{2-\beta}}.
\end{equation}

For counts with the observed near--Euclidean slope of $\beta=1.5$,
this raises the probability to $P(A>3)=2.3\tau\simeq 0.2$. 
The probability of extreme amplification is lower, but here a
power-law approximation to the counts is inadequate. If we adopt
the count model from equation (25) of Peacock et al. (2000), 
we obtain a biased probability $P(A>30) \simeq 0.006$. 
Together with the higher velocity
dispersion needed for 3-586.0 in the on-axis case, this
strengthens the case for single-image lensing with a
moderate amplification $A\simeq 3$ in the case of HDF850.1.
The observation of such an amplification in one of the
first bright sub-mm sources from a blank-field survey is
consistent with our simple probability calculations,
which indicate that a substantial fraction (perhaps of order 1/5th)
of the high-redshift SCUBA sources
should be affected by lensing amplifications of this order
(Blain 1996; Blain et al.\ 1999).

It should be possible to test this prediction once deep radio and infrared 
follow-up imaging of large, unbiased  blank-field SCUBA surveys 
(such as the `8-mJy' survey; Scott et al. 2002) has been completed. For now
the best we can say is that a lensed fraction of this order seems not 
unreasonable. There is good evidence that, in a few cases,
SCUBA sources have been significantly lensed by an intervening 
galaxy $\simeq 1$--2 arcsec distant from the true identification,
based on radio or $K$-band imaging
(Smail et al.\ 1999; Frayer et al. 2000). However, in at least
one of these cases the lensing galaxy is associated with a 
foreground cluster used to boost the sensitivity of the SCUBA survey
(Smail et al.\ 1999) and thus 
the likelihood of such a lensing system may well be substantially 
greater than in blank field SCUBA surveys.

Indeed, in contrast to HDF850.1, none of the three other 
best-studied SCUBA sources 
selected from unbiased,
blank-field surveys shows any evidence of having been lensed 
by an intervening galaxy (CUDSS14A -- Gear et al. 2000; Lockman850.1 -- Lutz
et al. 2001; ELAISN2850.2 -- Dunlop 2001a). Moreover, this is also 
true for the 3 bright mm sources recently discussed by Dannerbauer
et al. (2002). It thus seems clear that significant 
lensing is {\it not\/} in general responsible for the bright sub-mm/mm source
population. On the other hand, lensing of a significant minority of the
sources detected in the `8-mJy' survey may well offer an explanation
for the apparent correlation between the SCUBA sources, and galaxies 
at relatively modest redshift found in the ELAIS N2 field by
Almaini et al. (2002). 

\subsection{Cosmic star-formation history revisited}

We conclude this discussion by considering how the new information
we now possess on HDF850.1 impacts on the original estimation of 
dust-enshrouded star-formation density at high redshift performed
by Hughes et al. (1998). The three main issues to consider are
(i) the improved confidence in radio--FIR photometric redshifts which follows
from the discovery of HDF850.1K, (ii) revised constraints on the plausibility
that HDF850.1 and other SCUBA sources are powered by AGN rather than 
starbursts, and (iii) appropriate corrections for gravitational lensing in
the light of this study.

\subsubsection{Estimated redshifts}
One important consequence of the discovery of HDF850.1K, is that 
it provides renewed confidence that the high SCUBA-source redshifts 
inferred from radio--far-infrared SED fitting should be trusted, even 
when the SCUBA source itself appears (at first, or even second sight) 
to be associated with a galaxy of surprisingly modest redshift.
It also reinforces the importance of very 
deep radio, mm and infrared imaging for SCUBA source follow-up 
prior to attempting spectroscopy of candidate identifications
within the original SCUBA error circle (e.g. Barger et al. 1999b).

If one trusts the radio--far-infrared 
SED-based redshift estimates, then current 
evidence suggests a median redshift $z \simeq 3$ for the bright SCUBA
galaxy population (Dunlop 2001b; Smail et al. 2000), with the substantial
majority of SCUBA sources lying at $z > 2$. Equally, the relative ease with 
which several other bright SCUBA sources have been identified compared to
HDF850.1 suggests the majority probably lie at 
$z < 4$. Thus, it still seems reasonable to follow Hughes et al. (1998) 
and assume an approximate 
redshift range $2 < z < 4$ for calculating the contribution made by
sub-mm sources to star-formation density.
 
\subsubsection{X-ray constraints on AGN activity}

Evidence continues to grow that SCUBA and Chandra/XMM  sources, while perhaps 
correlated on large scales, are rarely coincident (e.g.
Fabian et al. 2000; Bautz et al. 2000; Hornschemeier et al. 
2000; Barger et al. 2001a,b; Almaini et al. 2002).
As already discussed in the introduction, for the case of HDF850.1, 
the nearest detected X-ray source lies $\simeq 5$\,arcsec south-west of 
the SCUBA source. However, the extreme depth of the 1\,Msec Chandra image 
of the HDF, coupled with the fact we now possess such an accurate position
for HDF850.1, makes it of interest to calculate what {\it limits} can be placed
on the presence of an AGN in this source.

We have deduced 3-$\sigma$ upper limits on the X-ray flux density of
HDF850.1 using the Bayesian method of Kraft, Burrow \& Nousek (1991).
This yields a limit of $< 4 \times 10^{-20}\,{\rm W\,m^{-2}}$ 
in the soft band ($0.5-2$keV), and  
$< 3 \times 10^{-19}\,{\rm W\,m^{-2}}$ in the hard band ($2-8$ keV)

The resulting submm--to--X-ray ratios yield a spectral index 
$\alpha_{sx} > 1.4$. Reference to Fig. 4 of Almaini et al. (2002) 
demonstrates that such a value is entirely consistent with a starburst,
but means that an AGN cannot be
powering the sub-mm emission unless the obscuring column is Compton--thick
with a negligible scattered fraction ($\ll 1$\%). 

What limits can we place on the luminosity/black hole mass of any AGN
present in HDF850.1? This clearly depends on the level of absorption assumed.
If we assume no absorption, than adopting  $z = 4$ yields a limiting 
X-ray luminosity of  $L < 3 \times 10^{36}\,{\rm W}$, comparable
to the output of a weak Seyfert galaxy. Using a standard
bolometric correction (a factor $\simeq 12$), the assumption of 
Eddington-limited accretion yields an upper limit on 
black-hole mass of $3 \times 10^6\,{\rm M_{\odot}}$.
                             
If we consider a range of absorbing columns, the upper limits on the
inferred X-ray luminosity and black-hole mass are as summarized in Table 4.
Thus it can be seen that unless the absorbing column is Compton-thick,
the limiting black-hole mass is $\simeq 10^7\,{\rm M_{\odot}}$.

\begin{table}
\caption{\small Inferred 3-$\sigma$ upper limits on 
X-ray luminosity and black-hole mass for HDF850.1, 
assuming a range of absorbing columns}
\begin{tabular}{|l|l|l|} \hline \hline
$n_{\rm H} / {\rm m^{-2}}$ & $L_X / {\rm W}$ & $M_{bh} / {\rm M_{\odot}}$\\ \hline 
$10^{27}$             &  $4\phantom{.0}  \times 10^{36}$&$4\phantom{.0} \times 10^6$\\
$10^{28}$             &  $6\phantom{.0}  \times 10^{36}$&$6\phantom{.0} \times 10^6$\\
$10^{29}$             &  $1.5            \times 10^{37}$&$1.5 \times 10^7$\\
Comp thick (1\% scat) &  $3\phantom{.0}  \times 10^{38}$&$ 3\phantom{.0} \times 10^8$\\ \hline \hline
\end{tabular}

\end{table}

\subsubsection{Impact of gravitational lensing}

Although the results of this study indicate that the flux density of 
HDF850.1 has almost certainly been boosted by gravitational lensing, the 
inferred magnification factor is relatively modest, $\simeq 2-3$.
There is also no evidence to support the existence of a similar lensing 
system for any of the other HDF SCUBA sources. This is not surprising since
it is the
brightest source uncovered that is statistically most likely to be 
significantly lensed, and the occurrence of one case of lensing within the 5 
sources reported by Hughes et al. (1998) is certainly consistent
with the estimates of lensing prevalence given above.

It therefore seems reasonable to simply scale down the intrinsic
sub-mm luminosity of HDF850.1 by a factor of 3, for the purpose of
re-calculating 
the total comoving star-formation density of the 5 SCUBA sources in the 
HDF. This produces a reduction of $\simeq$ 30\% in the estimate
of comoving star-formation density reported by Hughes et al. (1998).

\section{Conclusion}
This study has provided a particularly striking demonstration of the 
importance of ultra-deep near-infrared and radio imaging for the 
successful identification of even the brightest of sub-mm selected sources.
The contrast between the sub-mm and optical views of the HDF could hardly
be more striking, with the host galaxy of the brightest sub-mm source in 
this field transpiring 
to be one of faintest and reddest objects ever uncovered at 
near-infrared/optical wavelengths. At such imaging depths 
sub-arcsec astrometric accuracy is clearly crucial if an unambiguous 
identification is to be secured. 

It is obviously unrealistic to expect that multi-frequency data of 
this depth and quality will be easily achieved for large samples of bright
sub-mm sources. It is therefore worth briefly revisiting what would have 
been concluded about HDF850.1 without the enormous investment of observing
time made with the IRAM PdB interferometer, VLA+Merlin, and most recently
Subaru. As discussed at the beginning of this paper, on the basis of
statistical association one would have concluded that HDF850.1 was hosted
by the $z \simeq 1$ elliptical galaxy 3-586.0. Alternatively, given the
growing evidence that SCUBA source host galaxies are EROs, and given the 
uncertainty in raw SCUBA-source positions, another apparently plausible
(and statistically likely) conclusion would be that 
HDF850.1 is hosted by the nearby (5-arcsec distant) 
ERO which is also the nearest 
obvious Chandra and VLA source (VLA 3651+1221). 
The former error would lead to the 
conclusion that HDF850.1 lies at relatively modest redshift, and that 
sub-mm--radio SED-based redshift estimation cannot be trusted. The latter
error would lead to the conclusion that HDF850.1 is an AGN. 

In fact of course neither of these previously possible conclusions 
can now be viewed as tenable, and we find that HDF850.1 does indeed appear
to be a violently star-forming galaxy at $z \simeq 4$, which could never 
be discovered via Lyman-break selection techniques. 

\section*{ACKNOWLEDGEMENTS}
This work was
based in part on data collected at the Subaru Telescope, 
which is operated by the National Astronomical Observatory 
of Japan. MERLIN is a UK National Facility operated by the University of
Manchester at Jodrell Bank Observatory on behalf of PPARC.
This work was based in part on 
observations with the NASA/ESA {\it Hubble Space Telescope},
obtained at the Space Telescope Science Institute, which is operated by the 
Association of Universities for Research in Astronomy, Inc. under NASA 
contract No. NAS5-26555.
James Dunlop acknowledges the enhanced research time provided by the 
award of a PPARC Senior Fellowship. Ross McLure, Bob Mann and Graham Smith
also acknowledge the support of PPARC. Ian Smail and Omar Almaini 
acknowledge support from the Royal Society.
David Hughes and Itziar Aretxaga's
work is supported by CONACyT grants 32180-E and 23143-E.

\end{document}